\documentclass[11pt,superscriptaddress, reprint, pre, longbibliography]{revtex4-1}

\usepackage[utf8]{inputenc}
\usepackage{amsmath}
\usepackage[utf8]{inputenc} 
\usepackage[english]{babel}
\usepackage{tikz}
\usepackage{amsmath}
\usepackage{graphicx}
\usepackage[colorlinks=true, allcolors=blue]{hyperref}
\usepackage{amsfonts}
\newcommand{\Tau}{\tau}

\begin{document}

\author{Nathan Abitbol}
\email{nathan.abitbol1@universite-paris-saclay.fr}
\affiliation{
Universit\'e  Paris-Saclay, CNRS,  FAST,  91405,  Orsay, France
}

\author{Alex Hansen}
\affiliation{
PoreLab, Department of Physics,\\ Norwegian University of
Science and Technology, N-7491, Trondheim, Norway
}
\author{Alberto Rosso}
\affiliation{
Universit\'e  Paris-Saclay,  CNRS,  LPTMS,  91405,  Orsay, France
}

\author{Federico Lanza}
\affiliation{
PoreLab, Department of Physics,\\ University of Oslo, NO-0316, Oslo, Norway
}

\author{Laurent Talon}
\email{laurent.talon@universite-paris-saclay.fr}
\affiliation{
Universit\'e  Paris-Saclay, CNRS,  FAST,  91405,  Orsay, France
}

\begin{abstract}
We investigate pressure imposed two-phase flows in porous media where a Newtonian fluid displaces an immiscible Bingham yield-stress fluid. Using a pore-network model, we identify invasion patterns governed by three dimensionless parameters: the pressure-based capillary number $Ca_P$, the viscosity ratio $M$, and a number $R_\tau$, the ratio of yield stress to capillary forces. Beyond the classical Lenormand regimes (viscous fingering, capillary fingering, stable displacement), we discover two new regimes: a tree-like pattern at low viscosity ratios and a progressively widening column-like pattern at high viscosity ratios. We characterize all regime transitions and establish a breakthrough condition for the invading fluid.
\end{abstract}

\title{Newtonian fluid invasion into a Yield-Stress Fluid-Saturated Medium}

\maketitle

\section{Introduction}

The displacement of yield-stress fluids, fluids that require a minimum amount of stress to flow, in porous media in the presence of a second Newtonian phase, such as air or water, is encountered in a wide range of industrial and environmental applications.
Notable examples include enhanced oil recovery (EOR) through gas injection, where the mobilization of yield-stress fluids plays a critical role in improving recovery efficiency~\cite{ghannam12, sheng15}.
Similarly, soil consolidation processes often rely on the injection of cement or polymer slurries to stabilize porous structures~\cite{seiphoori22, zeighami22}.
Foam and clay suspensions are also used to create subsurface barriers for pollution containment~\cite{portois18} and to engineer dynamic seals in hydrogen storage reservoirs~\cite{abedi24}.
In the latter case, Laponite suspensions are employed due to their ability to transition from low-viscosity states during injection to high-viscosity, yield-stress states that effectively trap hydrogen~\cite{orujov26}.
These processes are governed by complex interactions between capillary, viscous, and yield-stress forces, which determine the flow patterns and displacement efficiency in porous media.
Understanding and controlling these mechanisms is essential for optimizing applications ranging from energy storage to environmental remediation.
This study aims to investigate the displacement of a yield-stress fluid by an immiscible Newtonian one.
More specifically, we aim to characterise the distinct flow patterns that appear depending on the properties of the fluids.

Two-phase flows involving Newtonian fluids in porous media have been the subject of extensive research~\cite{bear88,dullien91, sahimi95}.
In particular, since the seminal work of Zarcone and Lenormand~\cite{lenormand85}, it is now well established that distinct flow regimes emerge depending on the relative dominance of viscous or capillary effects.
Notably, they proposed a phase diagram that maps the various invasion patterns as a function of two key dimensionless numbers: the viscosity ratio and the capillary number.

Three distinct flow regimes have been identified in two-phase Newtonian displacements in porous media.
When viscous dissipation dominates and the invading fluid is significantly less viscous than the defending fluid, a viscous fingering regime emerges.
In this regime, the displacement front destabilizes, resulting in the formation of multiple preferential paths~\cite{saffman58,homsy87,sinha24}.
The resulting growth patterns have been shown to belong to the same universality class as diffusion-limited aggregation~\cite{maloy85,mathiesen06}.
Conversely, when the invading fluid is more viscous than the displaced fluid, the front remains stable, leading to a flat, compact displacement pattern~\cite{homsy87, cottin10, batenburg15}.
At low capillary numbers, capillary effects dominate, producing capillary fingering patterns that can be effectively described using invasion percolation models~\cite{xu98,dospital22,clement25}.
Recent studies have extended the classical Lenormand phase diagram to account for additional physical effects.
Primkulov \emph{et al.}~\cite{primkulov23} examined the influence of the contact angle at the fluid-solid interface on flow regime transitions.
Geistlinger \emph{et al.}~\cite{geistlinger24} experimentally characterized how heterogeneity and surface roughness affect flow behavior.

In this study, we propose extending the classical phase diagram to include visco-plastic effects for the invaded fluids.
Flows in porous media involving yield-stress fluids have received comparatively little attention in the literature.
For instance, in the single-phase case, the existence of a minimum pressure drop, $\Delta P_0$, required to initiate flow is a first consequence of the yield stress.
Furthermore, at low applied pressure drops, regions where the fluid remains at rest may coexist with flowing paths.
Due to the inherent disorder of the porous medium, each potential flow path has a different activation pressure drop. This results in a hierarchy of preferential flow channels~\cite{talon13a,talon13b,liu19, waisbord19, schimmenti23}.

The two-phase flow case has received even less attention than the single-phase scenario.
The most closely related study is that of Chen and Yortsos~\cite{chen05}, who employed pore-network simulations to investigate the invasion of a Newtonian fluid into a porous medium initially saturated with a Bingham fluid.
However, their analysis relied on two key assumptions. First, they considered the zero-viscosity limit for the invading fluid. Second, they neglected capillary forces.
Additionally, due to the complexity of the non-linear equation system and the computational constraints of the time, both the accessible system sizes and the range of physical variables explored were limited.
Despite these limitations, Chen and Yortsos identified a novel flow regime in which yield stress dominates viscous forces.
This regime is characterized by the formation of a single invading path, similar to the first flowing path observed in the single-phase case.
Consequently, they proposed a new region in the phase diagram; however, its boundaries remained poorly resolved due to the aforementioned limitations.
More recently, Pourzahedi and Frigaard~\cite{pourzahedi24} developed a pore-network model for a system closely related to the present study.
Their model incorporates the presence of liquid films and examines the influence of several key parameters, such as pore-size distribution and system size, on breakthrough time and displacement efficiency.
However, their approach neglects capillary forces and imposes a fixed flow direction within the network. The invading fluid is also assumed viscous-less, similarly to Chen and Yortsos.
Another contribution is the work of Li \emph{et al.}~\cite{li25}, who investigated invasion patterns using both experimental observations and two-dimensional Lattice Boltzmann simulations under mixed boundary conditions (imposed flow rate at the inlet and fixed pressure at the outlet).
Their findings demonstrate that the yield stress modifies the transition between invasion percolation and viscous fingering regimes.
Additionally, they observed flow patterns characterized by fewer branches and reduced tortuosity. They propose a phase diagram based on the sweeping efficiency, but do not report any new displacement regimes beyond those already identified in Newtonian fluids.
It is important to note, however, that their use of the full Stokes equation in the Lattice Boltzmann framework inherently limits the system size, the number of pores, and the range of physical parameters that can be explored.
We also note studies on the displacement of yield-stress fluids in Hele-Shaw cells, which can be analogous to porous media. These studies analyze experimentally and theoretically viscous fingering that can emerge when a yield-stress fluid is displaced by a Newtonian fluid \cite{coussot99,lindner00, eslami17}. The competition between capillary and yield forces has also been studied theoretically in \cite{lanza22}, by adopting a capillary bundle model for a two-phase non-Newtonian flow and assuming steady-state conditions, new relations for the flow rate-pressure drop characteristic were found.

To the best of our knowledge, a comprehensive understanding of the mechanisms governing invasion patterns in the presence of yield-stress fluids remains elusive, and clear criteria for distinguishing between the different flow regimes are still lacking.
The objective of this study is to extend the Lenormand phase diagram to incorporate yield stress behavior in the defending phase. To this end, we employ a dynamic pore-network model, based on the algorithm introduced in Ref.~\cite{sinha21}, where interfaces between the two phases are represented by menisci that influence the flow. Our results reveal that when yield-stress effects dominate, the system exhibits invasion patterns that deviate from the purely Newtonian case, with the morphology depending on both fluid viscosities and the imposed pressure. Specifically, we observe a tree-like pattern for low-viscosity invading fluids and a column-like pattern for high-viscosity invading fluids, both of which we characterize in detail as functions of the system parameters.
The paper is organized as follows. In Section~\ref{sec:model}, we present the pore-network model. Section~\ref{sec:low_viscosity} derives a breakthrough criterion and describes the flow regimes arising when the invading fluid is much less viscous than the defending fluid. Section~\ref{sec:high_viscosity} addresses the opposite limit, where the Newtonian invading fluid is much more viscous than the defending fluid. Finally, Section~\ref{sec:phase_diagram} summarizes our findings in the form of a phase diagram.

\section{Model/Numerical Method} \label{sec:model}

\begin{figure*}[ht]
    \centering
    \includegraphics[width=0.8\hsize]{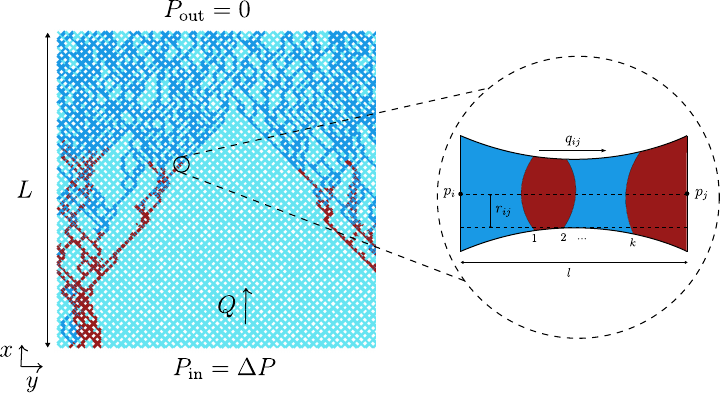}
    \caption{ Illustration of the pore-network model. The system consists of a diamond lattice of length $L$ and width $W$. The invading Newtonian fluid is shown in red, while the defending yield-stress fluid is shown in blue (light blue when immobile, dark blue when mobilized). Each link is assumed to have a hour-glass geometry, and can contain up to five menisci. A pressure $P_{\mathrm{in}} = \Delta P$ is imposed at the bottom boundary, while $P_{\mathrm{out}} = 0$ is imposed at the top boundary, driving flow from bottom to top.
    }
    \label{fig:sketch}
\end{figure*}

In this study, we investigate the invasion of a porous medium, initially saturated with a yield-stress fluid, by a less wetting Newtonian fluid. Our approach employs a pore network model, which simplifies the porous structure by representing pores as nodes and pore throats as links. Specifically, we consider a two-dimensional diamond network of width $W$ and length $L$, where each node $i$ is associated with a pressure $P_i$ (see Fig. \ref{fig:sketch}).

The Newtonian fluid is injected at the bottom boundary of the network. Each link $(i,j)$ is modeled as a hour-glass shaped conduit with a length $l$ and a radius $r_{ij}$. The radii are randomly sampled from a uniform distribution ranging from $r_{\mathrm{min}}=0.17$ to $r_{\mathrm{max}}=0.32$. A global pressure drop $\Delta P$ is applied along the longitudinal direction $L$, with $P_{\mathrm{in}} = \Delta P$ at the inlet nodes and $P_{\mathrm{out}} = 0$ at the outlet nodes. 
Mass conservation at each node $i$ is governed by the Kirchhoff condition:
\begin{equation}
    \sum_{j \in V(i)} q_{ij} = 0,
    \label{eq:khirchoff}
\end{equation}
where $V(i)$ represents the set of neighboring nodes of $i$.

For each link, the presence of two immiscible phases implies the formation of menisci between the yield-stress fluid and the invading Newtonian fluid (see Fig.~\ref{fig:sketch}). Both the menisci and the yield stress determine the onset flow condition in the link. A creeping  laminar-flow regime is assumed, for which inertial terms can be neglected. Following \cite{chen05,liu19,pourzahedi24}, the flow rate of the yield-stress fluid is modeled using a piecewise-linear approximation of Bingham rheology, characterized by a yield stress $\tau_c$. The flow rate -pressure relationship is given by:
\begin{equation}
    q_{ij} =
    \begin{cases}
        \kappa_{ij} \left( \delta P_{ij} - \Tau_{ij} - P^{\sigma}_{ij} \right), & \text{if } \delta P_{ij} - P^{\sigma}_{ij} > \Tau_{ij}, \\
        0, & \text{if } |\delta P_{ij} - P^{\sigma}_{ij}| \leq \Tau_{ij}, \\
        \kappa_{ij} \left( \delta P_{ij} + \Tau_{ij} - P^{\sigma}_{ij} \right), & \text{if } \delta P_{ij} - P^{\sigma}_{ij} < - \Tau_{ij},
    \end{cases}
    \label{eq:bingham_link_flow}
\end{equation}
where $\kappa_{ij} = (\pi/8) r_{ij}^4/(s_n\mu + (1-s_n)\eta)$ is the hydraulic conductivity of the link, where $s_n$ denotes the Newtonian saturation in the link, while $\mu$ and $\eta$ represent the plastic viscosities of the Newtonian fluid and the yield-stress fluid, respectively. The pressure difference is $\delta P_{ij} = P_i - P_j$, and the local yield-stress pressure threshold is given by:
\begin{equation}
    \Tau_{ij} = \frac{2 \tau_c l_{ij}}{r_{ij}}.
    \label{eq:tauij_def}
\end{equation}
The capillary pressure drop $P^{\sigma}_{ij}$ is determined from the location $x_k$ of every meniscus $k$. Following \cite{sinha21}, the link radius is assumed to be sinusoidal, thus for a link with $n$ menisci:
\begin{equation}
    P^{\sigma}_{ij} = \sum_{k=1}^n \pm \frac{2\sigma}{r_{ij}}
    \left[ 1 - a \cos\!\left(\frac{2\pi x_k}{l}\right) \right],
\end{equation}
where $\sigma = \hat{\sigma} |\cos\theta|$, with $\hat{\sigma}$ the surface tension between the two fluids and $\theta$ the contact angle between the capillary wall and the fluids interfaced, assumed to be constant in the whole network. The sign in the sum reflects the orientation of the meniscus. In our simulation, we used $a=1$ for the link geometry. Eq.~\eqref{eq:tauij_def} assumes a straight-tube geometry. However, for sinusoidal or more complex geometries \cite{roustaei13,roustaei16}, the corresponding pressure threshold differs only by a constant prefactor. This factor can be absorbed into a rescaling of the yield stress $\tau_c$ without loss of generality.

The algorithm proceeds through the following three steps.
First, the pressure and flow fields are computed to satisfy Eq.~\eqref{eq:khirchoff} and Eq.~\eqref{eq:bingham_link_flow} using the Augmented Lagrangian method, as outlined in~\cite{talon20}.
The velocity in each link is then used to displace the menisci.
The time step, $\delta t$, is determined based on the largest meniscus speed, $v_{\mathrm{max}} = \max_{ij} q_{ij}/\pi r_{ij}^2 $, which is calculated as follows:
$ \delta t = 0.1l / v_{\mathrm{max}}$.
Subsequently, in each node, the volume of wetting and non-wetting fluid from the ingoing links is redistributed proportionally to the flow rates of the outgoing links, following the algorithm described in~\cite{sinha21}. 
Finally, menisci that are close to each other (at a distance less than $10^{-4} l$) are merged, and a maximum number of menisci (five in our simulations) is enforced.

In the following, we denote average quantities by $\langle . \rangle$. We nondimensionalize all quantities using the characteristic length $l$, the characteristic (yield-stress) pressure threshold $\langle \tau \rangle = 2\tau_c/\langle r \rangle$, and the yield-stress fluid viscosity $\eta$. This leads to the introduction of several dimensionless numbers:

\begin{itemize}
    \item The viscosity ratio, defined as the ratio of the Newtonian viscosity to the yield-stress (plastic) viscosity:
    \begin{equation}
        M = \frac{\mu}{\eta}.
        \label{eq:M_def}
    \end{equation}

    \item The pressure-imposed Bingham number:
    \begin{equation}
        Bn_P = \frac{L \langle \tau \rangle}{\Delta P},
        \label{eq:Bn_def}
    \end{equation}

    \item The ratio of the yield stress to the capillary forces:
    \begin{equation}
        R_\tau = \frac{L \langle \tau \rangle}{2\sigma / \langle r \rangle} = \frac{L \tau_c}{\sigma},
        \label{eq:Rtau_def}
    \end{equation}
\end{itemize}

For convenience, we also define a capillary number. However, unlike the standard capillary number, which is based on the total flow rate imposed at the inlet \cite{chen05}, we define it in terms of the imposed pressure difference:
\begin{equation}
    Ca_P = \frac{\Delta P - \Delta P_0}{2\sigma / \langle r \rangle},
    \label{eq:Ca_def}
\end{equation}
where $\Delta P_0$ is the minimal pressure difference required to initiate flow.

As discussed in the Introduction, the present model is a continuation of the works of Chen and Yortsos \cite{chen05}, and Pourzahedi and Frigaard \cite{pourzahedi24}, which describe the two-phase flow invasion of a Newtonian fluid into a medium saturated with a yield-stress fluid.
We note that, although both studies neglect capillary forces, Pourzahedi and Frigaard account for the deposition of films of Bingham fluid on the throat walls when displaced by invading gas. In contrast, our model includes capillary forces at the menisci for partially wetting fluids, but does not account for liquid films.
Furthermore, in contrast to the latter work, the flow in our model is not constrained to be directed, a factor which can significantly impact the flow properties, as previously emphasized in \cite{abitbol26}.

\section{Low viscosity ratio} \label{sec:low_viscosity}

We first consider the limiting case where the viscosity of the invading Newtonian fluid is much lower than that of the defending yield-stress fluid ($\mu \ll \eta$).

\subsection{Critical pressure for breakthrough}

The first step in determining the flow regimes is to establish the conditions for flow and breakthrough. This scenario differs from both the single-phase yield-stress case and the two-phase Newtonian case.

In the case of single-phase flow with a yield-stress fluid, the critical pressure corresponds to the onset of the first flowing path. This is determined by minimizing the sum of local pressure thresholds over all possible paths~\cite{roux87,talon13b}.

By contrast, for immiscible fluids, the breakthrough condition is governed by a percolation criterion due to capillary forces~\cite{hansen87}. This approach involves identifying the first breakthrough path, where each path becomes accessible once the applied pressure exceeds the largest capillary barrier along that path.

To incorporate both effects, we propose the following argument to determine the minimum breakthrough pressure. We observe that, for a link in the network to be blocked by both capillarity and yield stress, it must first be connected to the inlet by the Newtonian fluid.

\begin{figure}[h!]
    \centering
    \includegraphics[width=0.6\hsize]{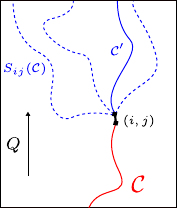}
    \caption{Illustration of the procedure for searching for the optimal path, see Eq.~\eqref{eq:DP0_def}. When a path of Newtonian fluid reaches link $(i,j)$, it needs to overcome both the local capillary barrier and the remaining yield stress constraint (path $\mathcal{C}'$).
    }
    \label{fig:sketch_minmax}
\end{figure}

Consequently, the key idea is to identify, for each possible flow path $\mathcal{C}$, the most restrictive (or blocking) link due to the combined effects of capillarity and the yield stress fluid. Specifically, if the invasion front has reached the link $(i,j)$, the minimal condition for further advancement corresponds to an imposed pressure:
\begin{equation}
    c_{ij} = \frac{2(1 + a)\sigma}{r_{ij}} + \min_{\mathcal{C'} \in \mathcal{S}_{ij}(\mathcal{C})} \sum_{(k,l) \in \mathcal{C'}} \tau_{kl}.
\end{equation}
Here, $\mathcal{S}_{ij}(\mathcal{C})$ represents the set of all possible flow paths in the yield-stress fluid connecting $(i,j)$ to the outlet (illustrated as the blue paths in Fig.~\ref{fig:sketch_minmax}), excluding any overlap with the already invaded path. For this path, the invasion by the Newtonian fluid proceeds iteratively from the inlet.

For each path $\mathcal{C}$, the critical pressure is identified as the maximum value of $c_{ij}$ along that path. The first path to open is then determined as the one with the smallest critical pressure among all possible paths. This minimal critical pressure, $\Delta P_0$, is thus defined as:
\begin{equation}
    \Delta P_0 = \min_{\mathcal{C}} \max_{(i,j) \in \mathcal{C}} \left[ \frac{2(1+a)\sigma}{r_{ij}} + \min_{\mathcal{C'} \in \mathcal{S}_{ij}(\mathcal{C})} \sum_{(k,l) \in \mathcal{C'}} \tau_{kl} \right].
    \label{eq:DP0_def}
\end{equation}

To compute $\Delta P_0$, we employed an adapted version of the min-max algorithm proposed in \cite{hansen87}, with the key modification that, at each iteration, the optimal path from a given link $(i,j)$ to the outlet is determined using Dijkstra's algorithm \cite{dijkstra59} while explicitly excluding the previously explored optimal path.

We observe that this formula accurately accounts for the two expected limiting cases.
First, if both fluids are Newtonian (i.e., $\tau_{kl} = 0$), the expression reduces to the min-max percolation criterion~\cite{hansen87}.
Second, if capillary effects are neglected, the maximum threshold along any path will necessarily occur at the first link (connected to the inlet), since all $\tau_{kl}$ are positive.
As a result, the selected path coincides with that of the single yield-stress fluid case, and the critical pressure simplifies to $  \Delta P^{\tau}_0 = \min_{\mathcal{C}} \left[ \sum_{(k,l) \in \mathcal{C}} \tau_{kl} \right]$. For example, we recover the MTP expression proposed by Chen and Yortsos \cite{chen05}, since they neglected capillary forces.

\begin{figure*}[ht!]
    \centering
    \includegraphics[width=0.35\hsize]{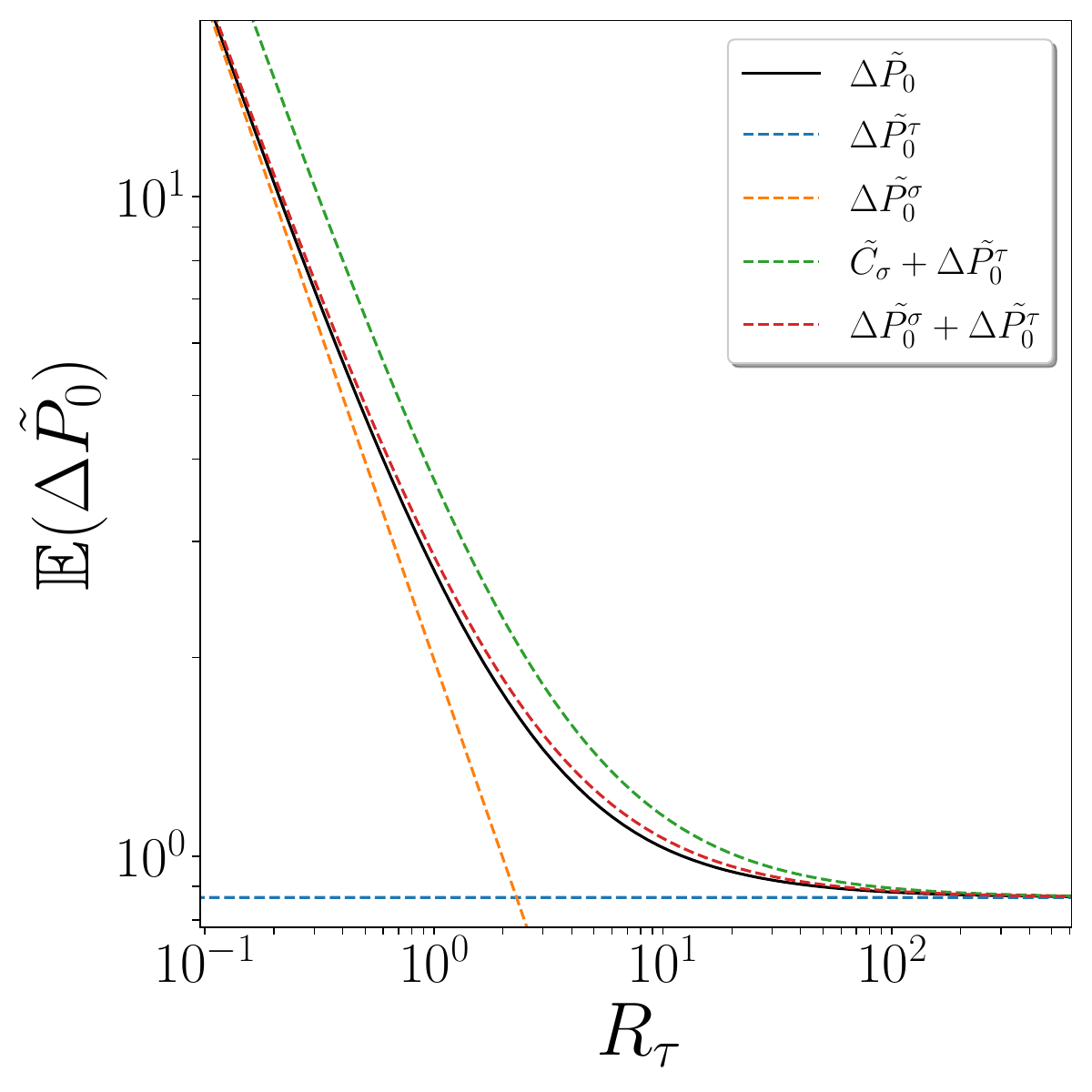}
    \includegraphics[width=0.35\hsize]{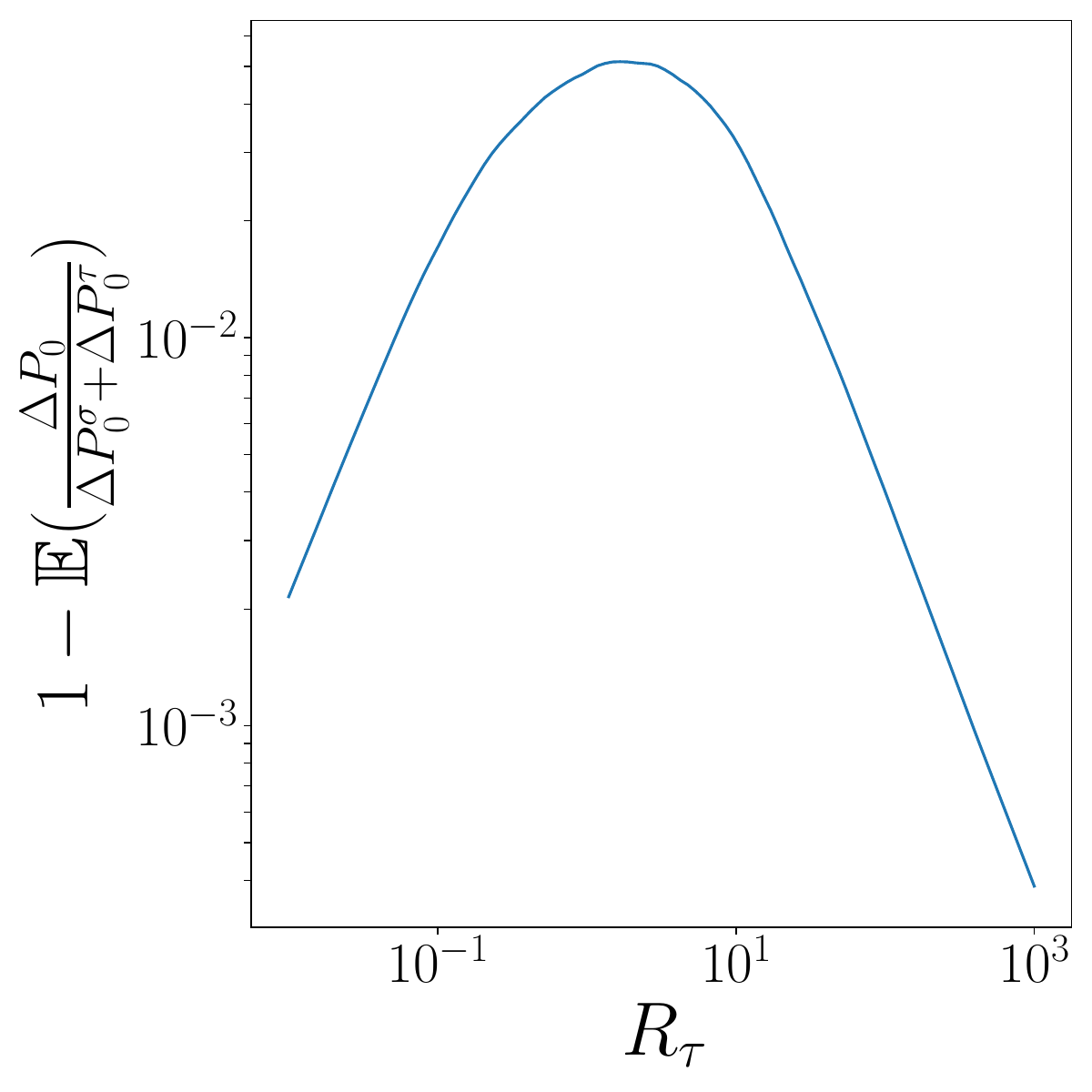}
    \caption{\textbf{Left:}
    Average over 100 realizations of the normalized critical pressure threshold, $\Delta \tilde{P}_0 = \Delta P_0/L\langle \tau \rangle$, as a function of $R_\tau$ for a system of size $L = 100$.
    \textbf{Right:}
    Relative difference between the exact value of $\Delta P_0$, computed using Eq.~\eqref{eq:DP0_def}, and the upper bound $\Delta P_0^\sigma + \Delta P_0^\tau$. Note that $\Delta {P_0^\tau} < L \langle \tau \rangle$ due the minimization over all possible paths. 
    }
    \label{fig:critical_pressure}
\end{figure*}

In Fig.~\ref{fig:critical_pressure} (left), we illustrate the evolution of the non-dimensional critical pressure, $\Delta P_0/L \langle \tau \rangle$, as a function of $R_{\tau}$. Two distinct asymptotic regimes emerge:
\begin{itemize}
    \item When $R_{\tau} \to 0$, the critical pressure converges to the percolation threshold, $\Delta P^{\sigma}_0$, which is independent of $\langle \tau \rangle$ and scales with $\sigma$ (orange line).
    \item Conversely, as $R_{\tau} \to \infty$, the critical pressure approaches $\Delta P^{\tau}_0$, a value proportional to $\langle \tau \rangle$ (blue line).
\end{itemize}

We note that the formulation given in Eq.~\eqref{eq:DP0_def} is computationally intensive for large systems.
However, an upper bound for the critical pressure can be derived by decoupling the two contributions:

{\begin{align}
    \Delta P_0  &\leq \min_{\mathcal{C}} \left[ \max_{(i,j) \in \mathcal{C}} \left( \frac{2(1+a)\sigma}{r_{ij}} \right) + \sum_{(k,l) \in \mathcal{C}} \tau_{kl} \right] \nonumber \\ & \leq  \min_{\mathcal{C}} \left[ \max_{(i,j)} \left( \frac{2(1+a)\sigma}{r_{ij}} \right) + \sum_{(k,l) \in \mathcal{C}} \tau_{kl} \right] \nonumber \\
    & = C_\sigma + \Delta P_0^{\tau},
\end{align}}

with $C_\sigma = \max_{(i,j)} \left( \frac{2(1+a)\sigma}{r_{ij}} \right)$. In Fig.~\ref{fig:critical_pressure}, we plot this upper bound (green line) along with $\Delta P_0^\tau + \Delta P_0^\sigma$ (red line) and observe that the latter quantity provides a very good approximation of the true critical pressure $\Delta P_0$, with an error of less than 6\% for the distribution shown (Fig.~\ref{fig:critical_pressure}, right). 
We emphasize that this result applies to a relatively weakly disordered medium; we expect the approximation error to increase for systems with higher heterogeneity.

The critical pressure threshold $\Delta P_0$ defines a region in the $(Bn_P, R_\tau)$ plane where breakthrough cannot occur. This prediction is confirmed numerically in Appendix~\ref{sec:DP0_check}.

\subsection{directed tree-like regime}

\begin{figure*}[ht!]
    \centering
    \includegraphics[width=0.8\hsize]{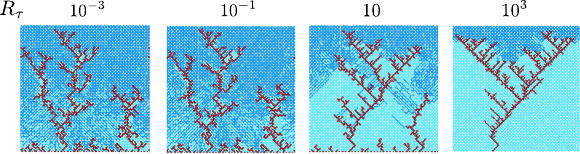}
    \caption{ Snapshots of simulations on a network with $Ca_P =0.07$ and system size $L = 80$, taken at breakthrough for multiple $R_\tau$ values. For low capillary numbers, the displacement shows a viscous fingering pattern. When yield stress effects become significant, the invasion takes the form of a directed, tree-like pattern.
    }
    \label{fig:tree_snapshots}
\end{figure*}

We investigate fluid displacement in the low viscosity ratio limit for pressures exceeding the critical threshold.
Fig. \ref{fig:tree_snapshots} presents invasion patterns observed at a fixed low capillary number for varying values of $R_\tau$.
As $R_\tau$ increases, the flow morphology evolves significantly.

At low values of $R_\tau$, the patterns closely resemble classical viscous fingering, which is consistent with observations in Newtonian fluids at low viscosity ratios (see, for example, \citep{lenormand88}).
In this regime, multiple heterogeneous fingers develop, with one dominant finger inhibiting the growth of the others.

As $R_\tau$ increases, the flow undergoes a transition towards a tree-like structure characterised by a primary invasion path.
This path progressively branches as it advances from the inlet, with the branches appearing straighter and more directed towards the outlet than in the viscous fingering regime.
At high $R_\tau$, a significant proportion of the displaced fluid near the inlet remains immobile and is trapped below the yield stress threshold. \newline

\paragraph*{\textbf{Tortuosity:}}

To quantitatively distinguish between these regimes, we propose measuring the \emph{directness} of the invasion tree by evaluating the tortuosity of the shortest path from the leaves to the root.
This tortuosity is determined using a breadth-first search algorithm and is defined as the ratio of the shortest path length from the tip to the root, $\ell_{\mathrm{tip} \to \mathrm{root}}$, to the vertical height of the tip, $h_{\mathrm{tip}}$.

\begin{figure}[ht!]
    \centering
    \includegraphics[width=0.8\hsize]{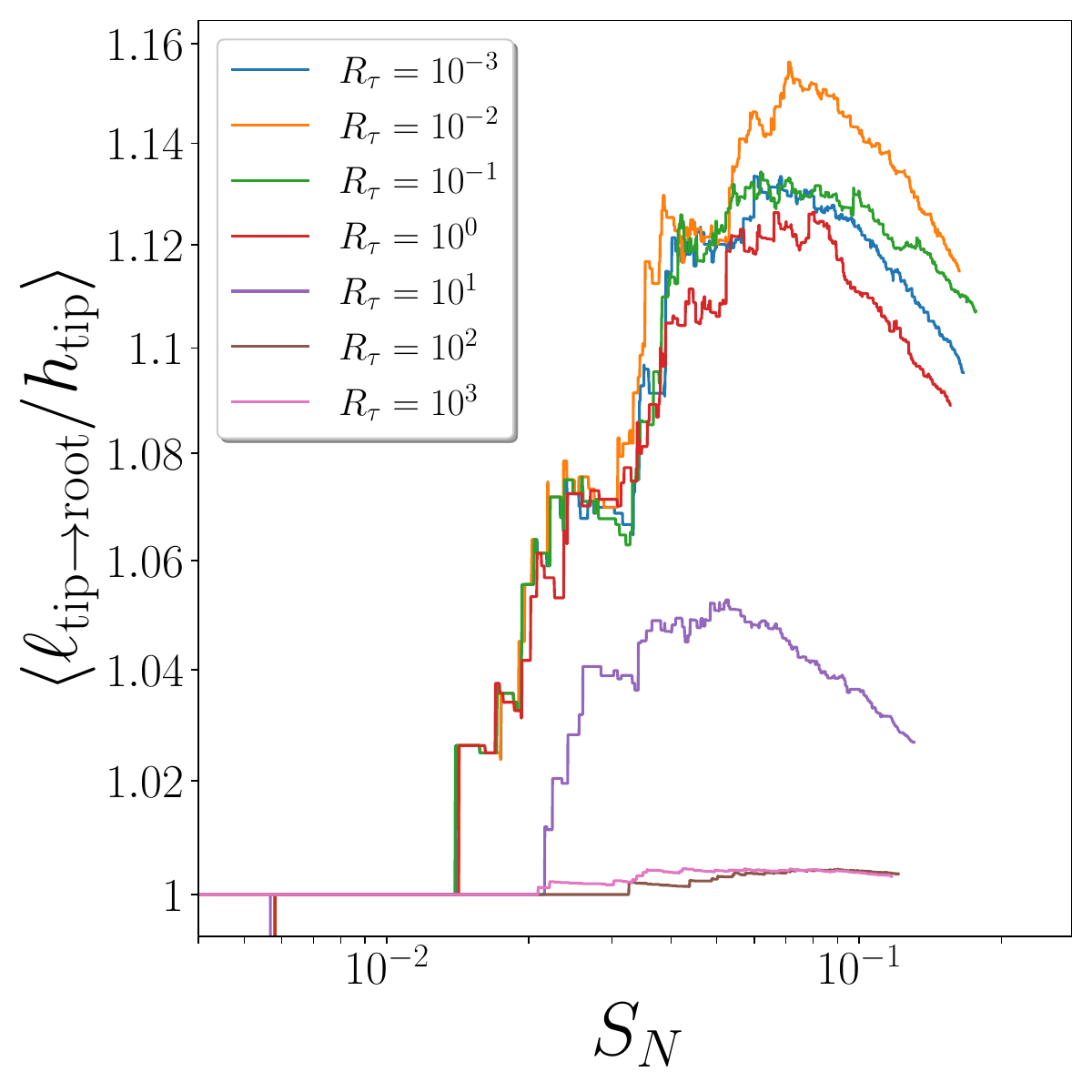}
    \caption{Average tortuosity of the Newtonian fluid paths connecting each tip of the invading fluid to the inlet as a function of Newtonian saturation $S_N$, for $Ca_P = 0.07$ and various values of $R_\tau$. The system size is $L = 80$. 
    }
    \label{fig:tortuosity_invasion_paths}
\end{figure}

In Fig.~\ref{fig:tortuosity_invasion_paths}, we plot the evolution of the average branch tortuosity as a function of saturation for various values of $R_\tau$.
For all cases, the tortuosity initially starts near unity but evolves differently depending on $R_\tau$.
At low values of $R_\tau$, the tortuosity increases continuously with saturation, reflecting the complex branching typical of viscous fingering.
In contrast, at high values of $R_\tau$, the tortuosity remains close to unity, indicating straighter, more directed invasion paths. We note that this tortuosity property of the invasion pattern is used to distinguish the two phases in the diagram in Section \ref{sec:phase_diagram}.
We observe that selecting the shortest path makes the tree's morphology dependent on the structure of the underlying network. In our network, for example, the tree takes the form of a diamond.

\begin{figure*}[ht!]
    \centering
    \includegraphics[width=0.8\hsize]{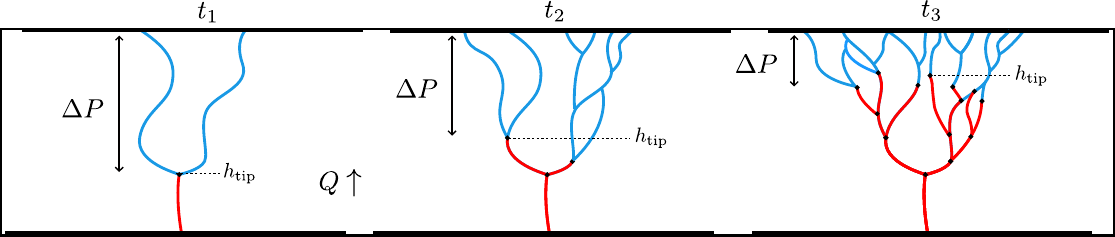}
    \caption{Illustration of the branching process typical of the directed tree regime, for three different time steps $t_1 < t_2 < t_3$. The mobilized yield stress fluid is represented is blue, the Newtonian fluid in red. As the Newtonian fluid invades the medium, the pressure is conserved in the fingers but the yield stress to overcome is reduced. The pressure gradient thus increases, causing more paths to flow in the defending phase.
    }
    \label{fig:tree_sketch}
\end{figure*}

\paragraph*{\textbf{Qualitative mechanism:}} 
To gain insight into the structure of the branching pattern, qualitative reasoning can be employed, as illustrated schematically in Fig.~\ref{fig:tree_sketch}. Assuming an extremely low viscosity ratio, i.e. that the viscosity of the Newtonian fluid is negligible, the pressure within the Newtonian fluid remains constant and equal to the imposed inlet pressure. The progress of the Newtonian fluid requires flow paths to exist within the yield-stress fluid. 
In a disordered medium, it is however known that the number of such flow paths depends on the difference between the mean pressure gradient and a threshold pressure gradient (see \cite{liu19}). 
Since the pressure is uniform in the Newtonian fluid, the number of paths is thus determined by the remaining distance between the tip and the outlet. Indeed, the mean pressure gradient in the yield-stress fluid scales as $\Delta P/(L - h_\mathrm{tip})$, where $h_\mathrm{tip}$ denotes the height of the tip.
As the front advances, the pressure gradient increases, thereby increasing the number of possible flow paths. 
In particular, at very low flow rates, the initial front consists of a single flow path. However, as $h_\mathrm{tip}$ increases, the number of flow paths grows, leading to the observed tree-like structure. Furthermore, we note that the flow condition in a yield-stress fluid strongly disfavors tortuous pathways. \newline

\begin{figure*}[ht!]
    \centering
    \includegraphics[width=0.3\hsize]{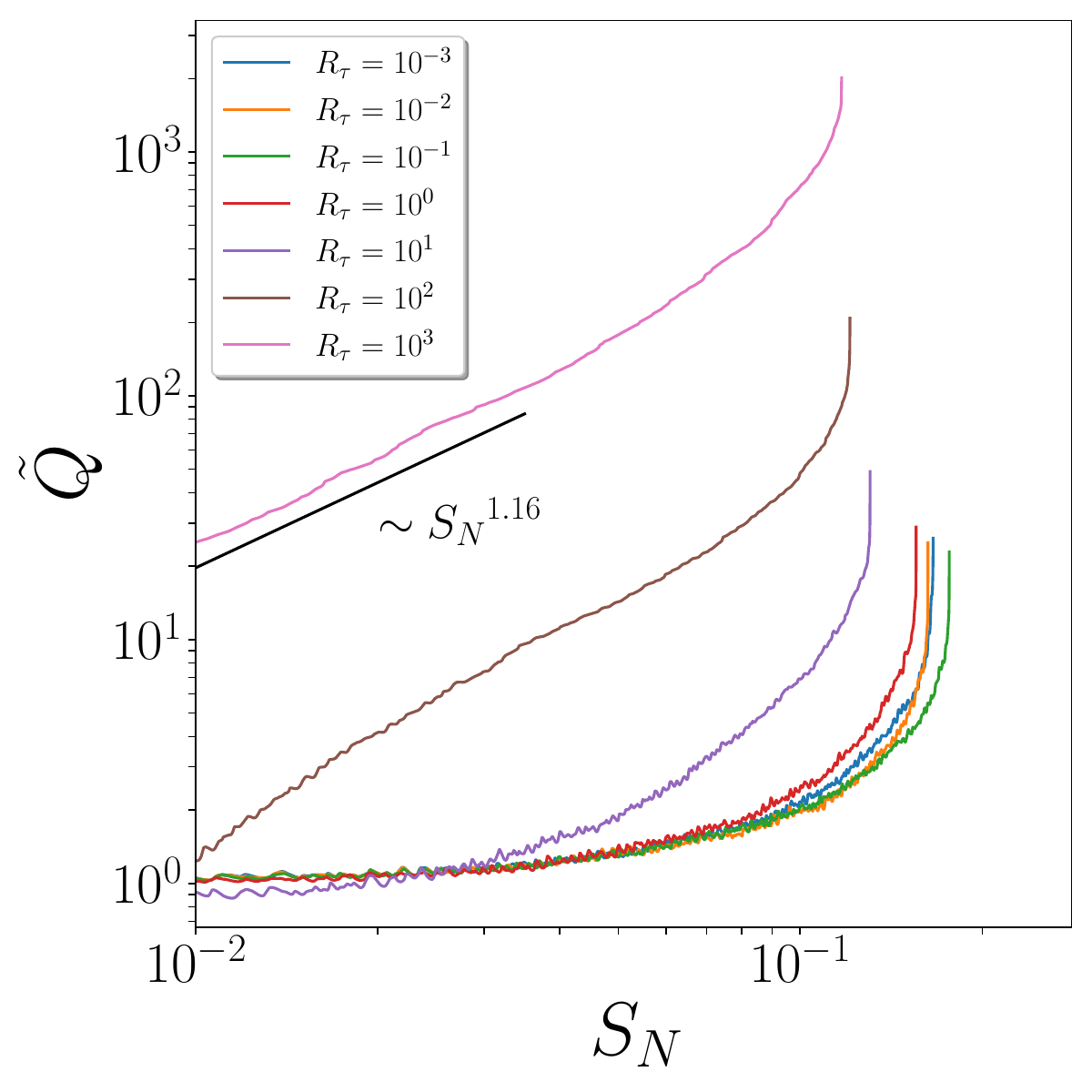}
    \includegraphics[width=0.3\hsize]{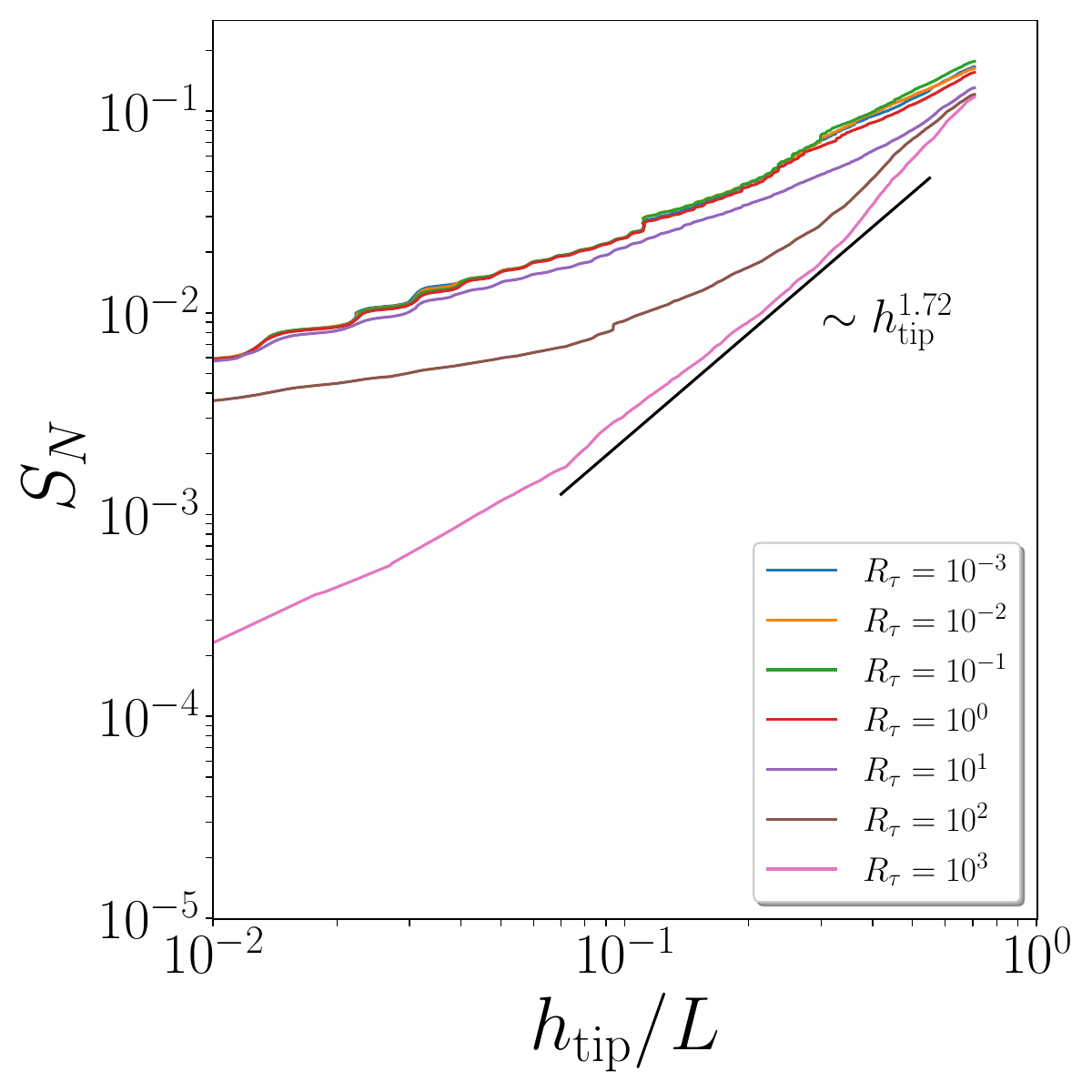}
    \includegraphics[width=0.3\hsize]{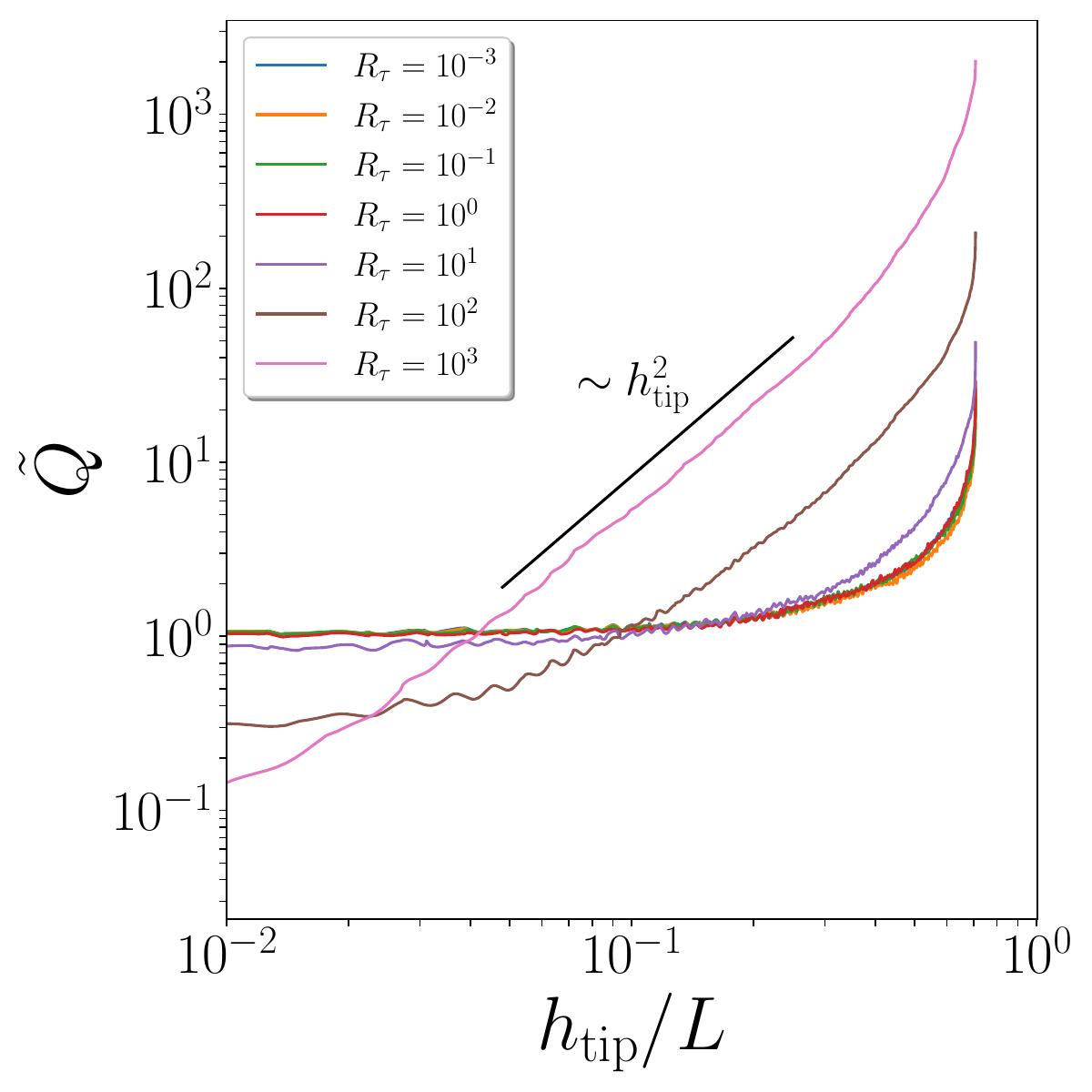}
    \caption{ \textbf{Left:} Normalized flow rate as a function of Newtonian saturation, for $M = 10^{-3}$, $Ca = 0.07$ and $L = 80$. When the yield stress effects dominate ($R_\tau \gg 1$), we observe the emergence of a power law behaviour : $Q \sim {S_N}^{1.16}$. This scaling is lost once capillary effects dominate. \textbf{Middle:} Saturation $S_N$ as a function of the most advanced tip's position $h_\mathrm{tip}$. A scaling behaviour $S_N \sim h_\mathrm{tip}^{1.72}$ is observed. \textbf{Right:} Normalized flow rate as a function of the tree height $h_\mathrm{tip}$.
    }
    \label{fig:flow_rate_scaling}
\end{figure*}

\paragraph*{\bf Flow rate-saturation relationship:} We now proceed by examining the temporal evolution of both the flow rate and the displacement efficiency during the invasion process. To this end, we first present in Fig.~\ref{fig:flow_rate_scaling} (left) the dimensionless flow rate as a function of saturation. The flow rate is non-dimensionalized by scaling with a characteristic flow rate \( Q_0 \), which corresponds to the typical flow of a homogeneous front in a homogeneous medium (the radii fluctuations are neglected) at very short times. Assuming that the flow initiates only when the pressure difference reaches \( \Delta P_0 \), it is written :
\begin{equation}
    Q_0 = \frac{W}{L}\frac{\pi \langle r \rangle^4}{8 \eta} (\Delta P - \Delta P_0)
\end{equation}
Our observations reveal that the evolution of the flow rate depends strongly on the value of $R_\tau$. For low values of $R_\tau$, the flow rate remains relatively modest during the early stages of the invasion and then increases gradually until breakthrough. A pronounced acceleration occurs just before breakthrough, which can be attributed to the divergence of the pressure gradient in the remaining defending fluid.

Conversely, for higher values of $R_\tau$, the flow rate increases from the onset of the invasion. The flow rate appears to scale with saturation according to a power law. Furthermore, at a given saturation, the flow rate is significantly higher than in cases where $R_\tau$ is much less than 1.
This behavior stems from the directed, tree-like structure of the flow: for instance, at high \( R_\tau \), the invasion initially proceeds along a single flow path, resulting in a more advanced tip height for a given saturation compared to the low \( R_\tau \) regime. Consequently, this leads to a higher pressure gradient in the remaining fluid domain. 

This behaviour is corroborated in Fig.~\ref{fig:flow_rate_scaling} (middle), which plots saturation as a function of the most advanced tip position.
Depending on the value of $R_\tau$, distinct trends are observed. For high values of $R_\tau$, the saturation follows a non-trivial power-law scaling with the height of the tree-like structure, $S_N \sim h_\mathrm{tip}^{D}$ with $D =1.72 \pm 0.03 $, suggesting a fractal morphology.

In Fig.~\ref{fig:flow_rate_scaling} (right), the evolution of the flow rate is presented as a function of the height of the most advanced tip. Due to the two scaling behaviours described above, the flow rate evolves differently in the high and low $R_\tau$ regimes. Specifically, the flow rate scales as $h_\mathrm{tip}^2$ at high $R_\tau$, whereas it remains nearly constant at low $R_\tau$ values.
To qualitatively account for the non-trivial power-law dependence of the flow rate on saturation, we propose a heuristic argument. We assume that the overall flow rate is governed by the flow within each individual tip, denoted as \( Q_\mathrm{tip} \). This quantity is determined by the pressure gradient in the yield-stress fluid over the remaining distance to be invaded. By taking into account both capillary forces and the remaining yield stress, it can be expressed as :
\begin{align}
Q_\mathrm{tip} & \approx \frac{1}{(L-h_\mathrm{tip}) \eta} \left( \Delta P - \frac{2 \sigma}{\langle r \rangle} - (L-h_\mathrm{tip}) \langle \tau \rangle \right) \nonumber \\
& \approx \frac{\langle \tau \rangle}{\eta (L-h_\mathrm{tip})} \left( \frac{LCa_P}{R_\tau} + h_\mathrm{tip} \right),
\end{align}
 where we introduced the dimensionless numbers defined in Eqs.~\eqref{eq:Ca_def} and \eqref{eq:Rtau_def} with the approximation $\Delta P_0 \approx L \langle \tau \rangle + 2\sigma/\langle r \rangle$.
Within a specific range of \( h_\mathrm{tip} \), such that \( Ca_P/R_\tau \ll h_\mathrm{tip}/L \ll 1 \) (early stage of the invasion, near the onset of mobilization), the flow rate per tip further reduces to
\begin{equation}
    Q_\mathrm{tip} \approx \frac{\langle \tau \rangle h_\mathrm{tip}}{\eta L}.    
    \label{eq:flow_rate_tip}
\end{equation}
Moreover, the tree-like structure of the system implies a proportionality between the number of tips and the height of the tree, \( n_\mathrm{tips} \propto h_\mathrm{tip} \). Assuming that all tips carrying flow are located at roughly the same height, and combining Eq.~\eqref{eq:flow_rate_tip} with the fractal scaling relation \( S_N \sim h_\mathrm{tip}^D \), we derive the scaling law for the total flow rate:
\begin{equation}
Q = n_\mathrm{tips} Q_\mathrm{tip} \sim S_N^{2/D}.
\end{equation}
The resulting exponent, \( 2/D \approx 1.16 \), is consistent with the value measured in Fig.~\ref{fig:flow_rate_scaling} (left). \newline

\begin{figure*}[ht!]
    \centering
    \includegraphics[width=0.8\hsize]{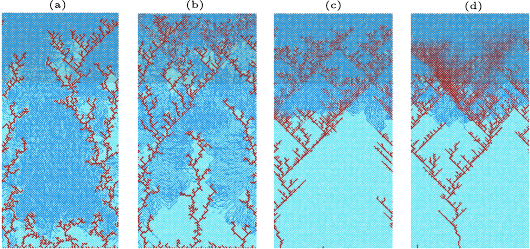}
    \caption{Snapshots of simulations performed on systems of size $L = 200$ and $W = 100$ with $R_\tau = 10^{-3}$ (a), $R_\tau = 10$ (b), $R_\tau = 10^{3}$ (c), and $R_\tau = 10^{4}$ (d).
    }
    \label{fig:larger_sizes_tree}
\end{figure*}

\paragraph*{\bf Foam transition:}
Similarly to the observations reported by Lanza \emph{et al.} \cite{lanza23} in pressure-driven two-phase flows, our experiments also suggest the emergence of a foam phase beyond a certain distance from the inlet. While a detailed investigation of this foam formation is beyond the scope of the present study, we note its occurrence. However, as illustrated in Fig.~\ref{fig:larger_sizes_tree}, the distance from the inlet at which foam appears seems to decrease with increasing \( R_\tau \).

\section{High viscosity ratio} \label{sec:high_viscosity}

\begin{figure*}[ht!]
    \centering
    \includegraphics[width=0.8\hsize]{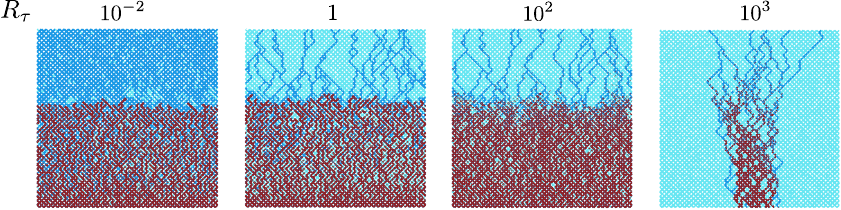}
    \caption{Snapshots of the simulation at breakthrough, obtained for $Ca_P =0.07$, $M = 10^3$ and $L = 80$. When the yield stress becomes dominant ($R_\tau \gg 1$) and the capillary number is sufficiently low, the displacement front destabilizes due to the formation of preferential flow paths. This process gives rise to a characteristic “column” invasion pattern.
    }
    \label{fig:snap_front_highM}
\end{figure*}

We now consider the complementary configuration, in which the invading Newtonian fluid has a higher viscosity than the yield-stress fluid (i.e. $M \gg 1$).

Figure~\ref{fig:snap_front_highM} shows the front displacement for a low capillary number $Ca_P$ and varying values of $R_\tau$.
At low $R_\tau$ and high $M$, the displacement exhibits the classical relatively stable Newtonian front, albeit with small regions of trapped defending fluid due to the presence of capillary forces.
As $R_\tau$ increases, the overall front displacement remains qualitatively similar, although the flow field upstream in the yield-stress region becomes increasingly channelised.
However, for sufficiently high values of $R_\tau$, the invasion pattern undergoes a radical transition: the invading fluid concentrates into a single column, leaving an entire column of yield-stress fluid at rest.

\subsection{Stable front}

In the stable front regime, although the displacement front appears qualitatively similar to the Newtonian case, in the presence of yield stress in the defending fluid, the dynamics of the front undergoes an important difference.

\begin{figure*}[ht]
    \centering
    \includegraphics[width=0.4\hsize]{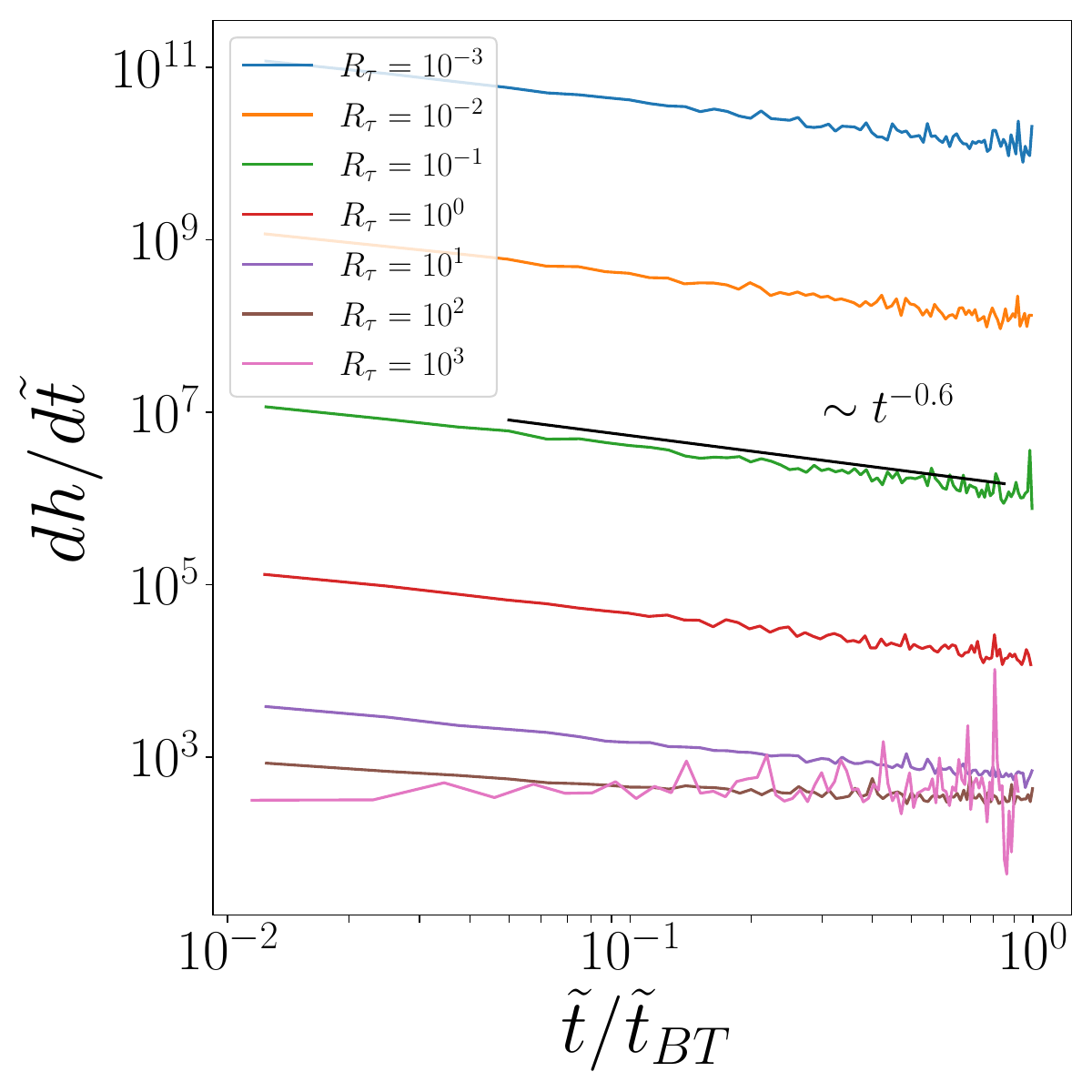}
    \includegraphics[width=0.4\hsize]{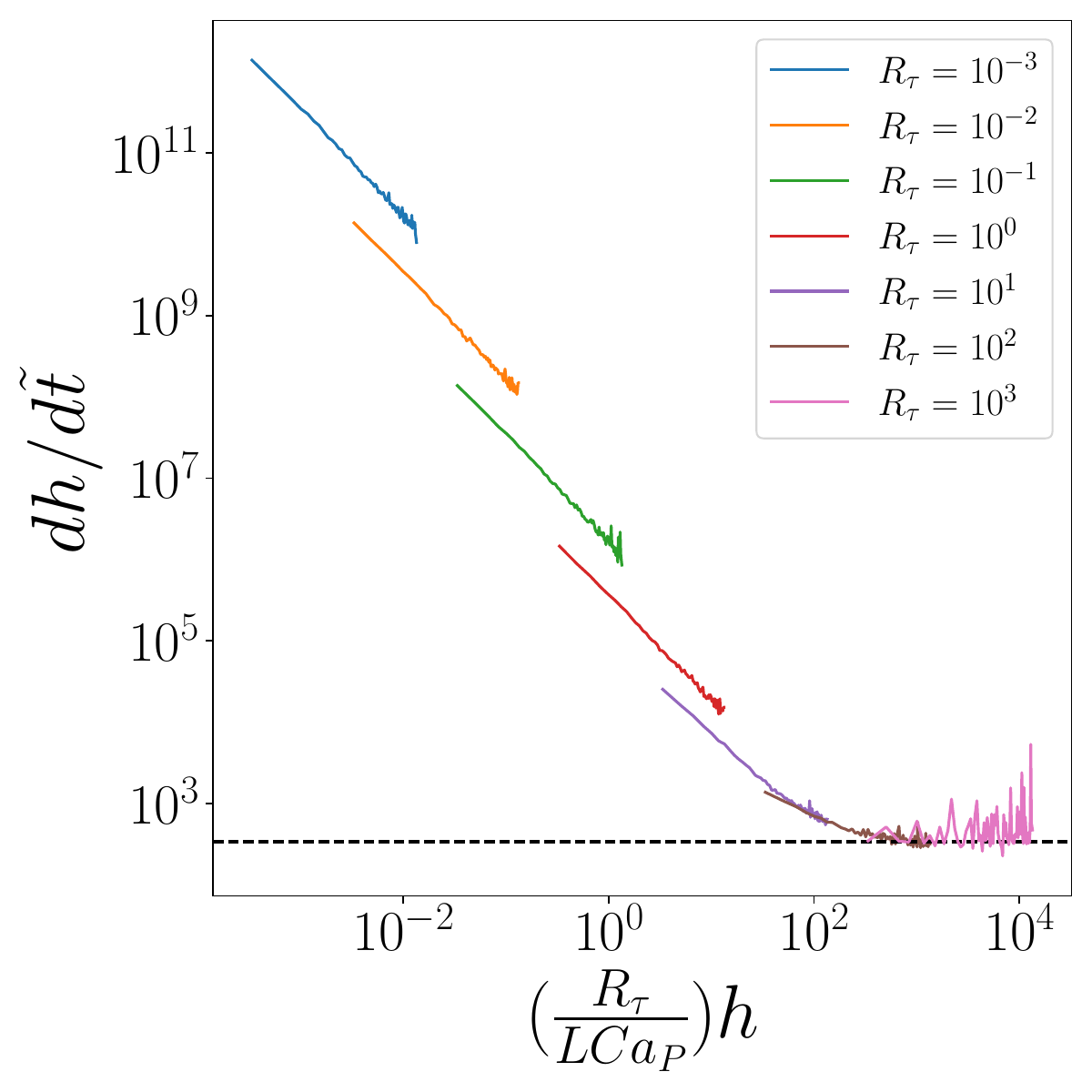}
    \caption{ Speed of the invasion front for $Ca_P = 0.07$ and various values of $R_\tau$, shown as a function of dimensionless time (left) and rescaled interface height (right). The dashed black line is the expected plateau value from Eq.~\eqref{eq:plateau_value}. The system size is $L = 80$. 
    }
    \label{fig:interface_speed}
\end{figure*}

Fig.~\ref{fig:interface_speed} (left) shows the front speed as a function of normalized time for various values of $R_\tau$.
At low $R_\tau$, the front velocity decays as a power law, $\propto t^{-0.6 \pm 0.02}$, which is reminiscent of the Washburn dynamics~\cite{washburn21} observed for Newtonian fluids.
In contrast, at higher $R_\tau$, the velocity decreases much more slowly, or even remains approximately constant.
The difference in dynamics is further confirmed in Fig.~\ref{fig:interface_speed} (right), which plots the front speed as a function of the front position, using a rescaling that will be discussed later.
At low $R_\tau$ or low $h$, the front speed follows a power-law decay, $\propto h^{-1.3}$, but it reaches a plateau at higher values of $R_\tau$.

\begin{figure}[ht]
    \centering
    \includegraphics[width=0.9\hsize]{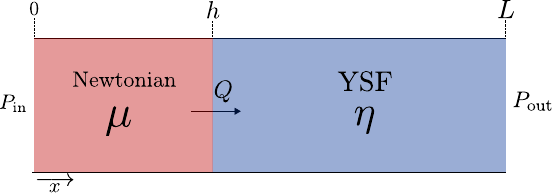}
    \caption{ Sketch of a homogeneous porous medium : the radii fluctuations are ignored. The Newtonian fluid of viscosity $\mu$ invades a medium of size $L$ initially saturated with a Bingham fluid of viscosity $\eta$. The interface between the two fluids is located at height $h$ and is assumed flat.
    }
    \label{fig:sketch_stable_front}
\end{figure}

To understand the evolution of the front speed, we adopt an approach analogous to Washburn's theory, but incorporate the effects of yield stress.
We consider a stable, compact invasion front progressing through a homogeneous medium, as illustrated in Fig.~\ref{fig:sketch_stable_front}.

Given the flatness and compactness of the invasion front, the interface speed is related to the flow rate by $Q = \pi \langle r \rangle^2 \frac{dh}{dt}$.
This leads to the following expression for the front velocity:
\begin{equation}
    \frac{dh}{dt} = \frac{\langle r \rangle^2/8\mu}{h + (L-h) M^{-1}} \left( \Delta P - (L-h)\langle \tau \rangle - \frac{2\sigma}{\langle r \rangle} \right).
\end{equation}

In the limit where the Newtonian fluid is highly viscous ($M = \mu/\eta \gg 1$) and the interface height $h$ is non-zero, the interface speed can be approximated as:
\begin{equation}
    \frac{dh}{dt} \approx \frac{\langle r \rangle^2}{8 \mu} \left( \Delta P - L \langle \tau \rangle - \frac{2\sigma}{\langle r \rangle} \right) \frac{1}{h} + \frac{\langle r \rangle^2}{8 \mu} \langle \tau \rangle.
    \label{eq:interface_speed_washburn}
\end{equation}

We define the effective pressure threshold as $\Delta P_\mathrm{eff} = L \langle \tau \rangle + 2 \sigma / \langle r \rangle$.
The front speed in Eq.~\eqref{eq:interface_speed_washburn} comprises two terms: a constant term and a term proportional to $1/h$, leading to two distinct regimes:
\begin{align}
    \frac{dh(t)}{dt} & \overset{t\rightarrow 0}{\propto} \frac{1}{h}, \\
    \frac{dh(t)}{dt} & \overset{t\rightarrow \infty}{\approx} \frac{\tau_c \langle r \rangle}{4 \mu}.
    \label{eq:plateau_value}
\end{align}
The constant speed regimes occurs when:
\begin{equation}
    h \gg \frac{\Delta P - \Delta P_\mathrm{eff}}{\langle \tau \rangle} \sim \frac{L \, Ca_P}{R_\tau}.
\end{equation}
The first regime corresponds to Washburn-like dynamics, as observed in Newtonian fluids, while the second regime is governed by the yield stress contribution.
Notably, the transition depends on the front height $h$, and for high $R_\tau$, it occurs closer to the inlet.

We observe that the power-law exponent ($0.6$) of the first regime in our results differ slightly from the theoretical prediction $1/2$, obtained by solving the ODE in Eq.~\eqref{eq:interface_speed_washburn} after setting $\langle \tau \rangle=0$. 
This discrepancy may arise from the non-compact nature of the displaced fluid. Indeed, a fractal structure would imply $Q \propto dh^{d_f}/{dt}$ with $d_f \neq 1$.
Nonetheless, this model reasonably captures the transition height and the plateau value observed in the simulations.

\subsection{Column regime}

As described above, for high $R_\tau$, the displacement front undergoes a significant change in morphology.
Instead of maintaining a relatively flat interface, the flow becomes highly channelized, with a single column of invading fluid advancing alongside a column of undisplaced fluid.

\begin{figure*}[ht]
    \centering
    \includegraphics[width=0.9\hsize]{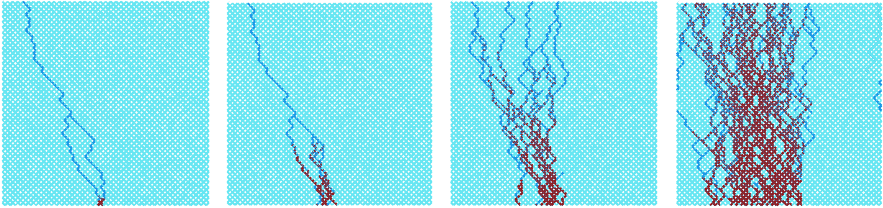}
    \caption{ Snapshots of the simulation at successive time steps for $Ca_P = 0.19$, $M = 10^3$ and $R_\tau = 10^4$. The system size is $L = 80$. Initially, a single Newtonian column penetrates the medium, after which secondary branches emerge, originating either from the inlet or from previously invaded paths. As a result, the invading column progressively widens over time.
    }
    \label{fig:snap_column}
\end{figure*}

To better understand the physical processes at work, we illustrate the temporal evolution of this regime in Fig.~\ref{fig:snap_column}.
At the very early stage, the invasion is confined to only a few links.
This behavior arises due to the limited number of flowing channels in the yield-stress fluid, a consequence of the combined effects of high $R_\tau$ and low $Ca_P$. As the displacement progresses, unlike in the tree-like or viscous fingering regimes, the column widens rather than splitting into multiple fingers.
This widening is driven by a branching process within the column, where new branches initiate either from the inlet or from the column itself.

\begin{figure}[ht!]
    \includegraphics[width=0.9\hsize]{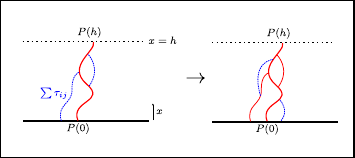}
    \caption{Sketch of a column formation and widening. A single channel of Newtonian fluid reaches height $h$, until the pressure drop $P(0) - P(h)$ is large enough to activate a new flowing path.}
    \label{fig:sketch_column}
\end{figure}

To qualitatively describe the branching process within the column (rather than tip splitting), one may adopt a reasoning analogous to that presented in Section \ref{sec:low_viscosity}.
In this case however, given that $M$ is large, the pressure gradient increases significantly along the column, therefore promoting the emergence of branching flow paths.
To illustrate this effect, we consider the pressure gradient along a single invasion path, as depicted in Figure~\ref{fig:sketch_column}.
Assuming a homogeneous medium, the pressure difference $\delta P(h)=P(0) - P(h)$ between the base and the tip of a single flow path is given by:
\begin{align}
\delta P(h) = \left(1 + \frac{L - h}{Mh}\right)^{-1}
\left( \Delta P - (L - h) \langle \tau \rangle - \frac{2\sigma}{\langle r \rangle} \right).
\label{eq:model_pressure_diff}
\end{align}
It is assumed here that, beyond the tip, the flow continues through a single mobilized channel within the Bingham fluid. From this expression, it is evident that the pressure gradient is negligible for small $h$, but increases as the invasion progresses thus increasing the probability for emergence of a new flowing branch.
A first approximation condition for the onset of branching in a yield-stress fluid can be estimated as 
\begin{equation}
    \delta P(h)/h \gtrsim \langle \tau \rangle
    \label{eq:branching_condition1}
\end{equation}.
This leads to a threshold distance for the emergence of new branches:
\begin{align}
h &\gtrsim L - M \frac{\Delta P - L \langle \tau \rangle - 2\sigma/\langle r \rangle}{\langle \tau \rangle} \nonumber \\
&\gtrsim L \left(1 - M \frac{Ca_P}{R_\tau}\right),
\end{align}
Notably, this reasoning also applies to any two points within the column.
However, we note that this prediction results in an overestimation of the branching height.
For instance, using the parameters of Fig. \ref{fig:snap_column}, it predicts that branching would begin only when $h > 0.98L$.
The reason for this discrepancy is that the condition given in Eq.~\eqref{eq:branching_condition1} is overly restrictive, as it neglects the optimization of the selected flow path.

Indeed, it is well established that the activation of new flow paths occurs below the average threshold, such that $\delta P(h)/h > c \langle \tau \rangle$, where $c < 1$~\cite{kardar87,liu19,abitbol26}.
This is due to the combined effects of disorder and path optimization.
Accounting for this correction, we derive the critical height for branching as:
\begin{equation}
    h_c = \left[1 - c\left(1 - \frac{1}{M}\right)\right]^{-1} \left( \frac{Lc}{M} - \frac{\Delta P - \Delta P_\mathrm{eff}}{\langle \tau \rangle} \right).
\end{equation}
This can be simplified to:
\begin{equation}
    \frac{h_c}{L} \approx \left[1 - c\left(1 - \frac{1}{M}\right)\right]^{-1} \left( \frac{c}{M} - \frac{Ca_P}{R_\tau} \right).
    \label{eq:height_branching}
\end{equation}

\begin{figure*}[ht!]
    \centering
    
    \includegraphics[width=0.4\hsize]{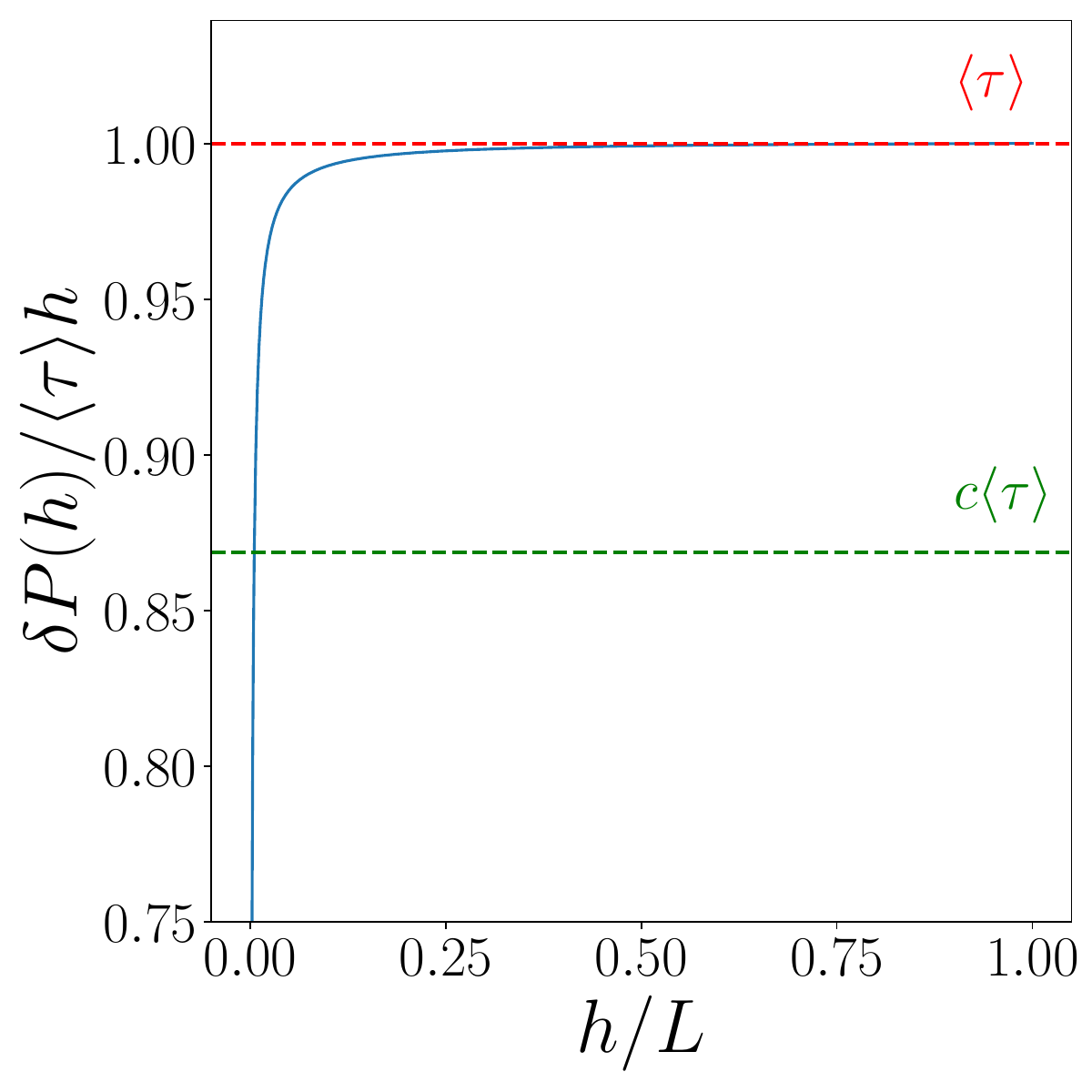}
    \includegraphics[width=0.4\hsize]{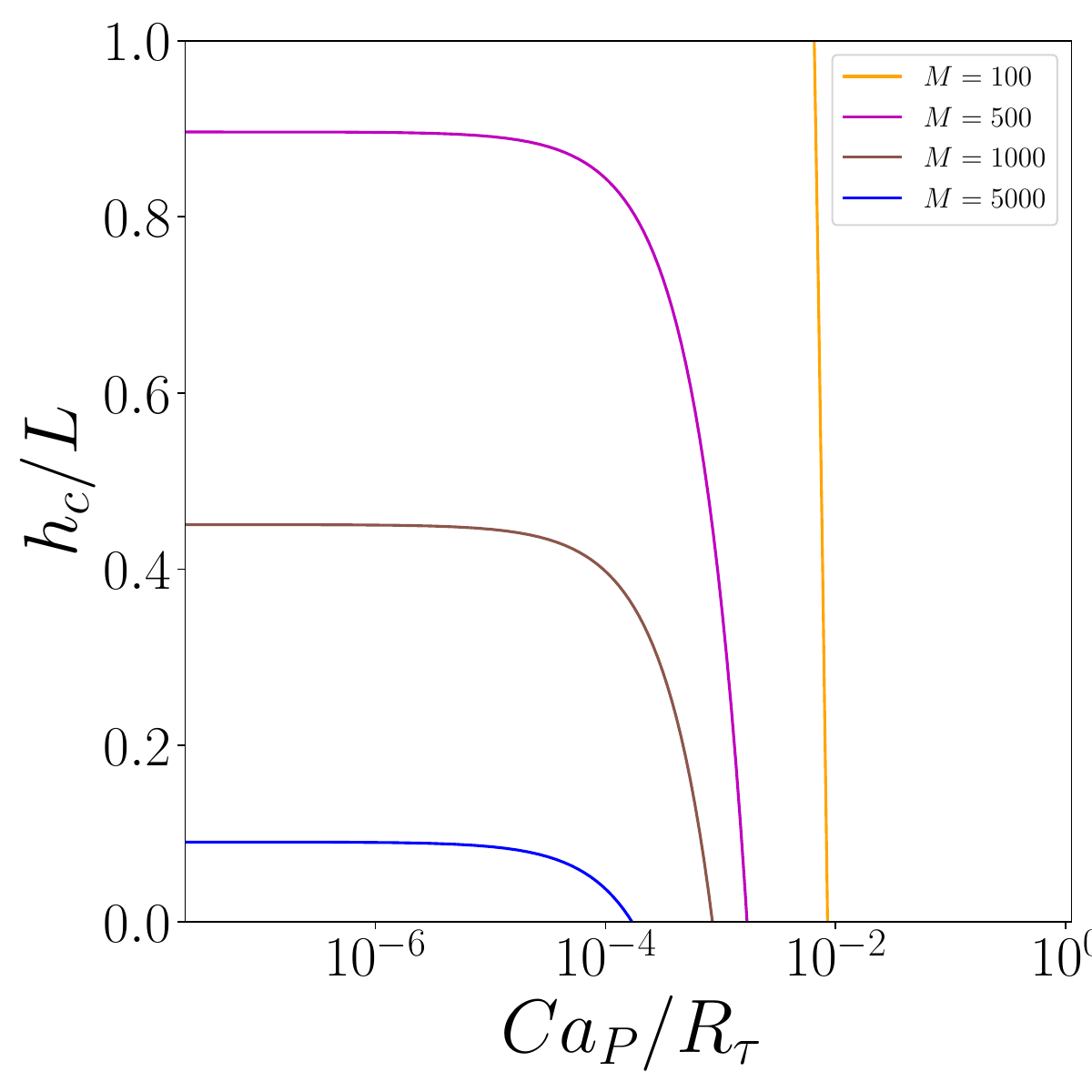}
    \caption{ \textbf{Left:} Normalized pressure gradient $\delta P(h)/h$ between the inlet and the tip of the invaded channel as a function of the tip's height $h$ from Eq. \eqref{eq:model_pressure_diff}, with $Ca_P = 0.2$, $R_\tau = 10^3$ and $M = 10^3$. The pressure gradient attains the values $c \langle \tau \rangle$ (green line) and $\langle \tau \rangle$ (red line) at radically different heights. \textbf{Right:} Height $h_c$ predicted for the branching processes, using Eq. \eqref{eq:height_branching}, as a function of the ratio $Ca_P/R_\tau$ and for multiple values of $M$. }
    \label{fig:pressure_gradient_column}
\end{figure*}

We emphasize that the position of $h_c$ is highly sensitive to the value of $c$, due to the very small slope of $\delta P(h)/h$ near $\langle \tau \rangle$, as illustrated in Fig.~\ref{fig:pressure_gradient_column} (left).
Using the value $c = 0.87$, measured for infinite system sizes (see also~\cite{abitbol26}), yields a significantly smaller $h_c$, in agreement with our observations.
In Fig.~\ref{fig:pressure_gradient_column}, we plot the value of $h_c$ using the estimation from Eq.~\eqref{eq:height_branching}.
We observe that the branching height converges to a constant value in the limit of low $Ca_P/R_\tau$ (high yield stress with respect viscous forces), where this constant increases with decreasing $M$.
This is consistent with the previous limit, where for very low $M$, no branching occurs in the invading flow path.
As $Ca_P/R_\tau$ increases, the value of $h_c$ decreases toward zero, implying that the advancement of a single link can trigger flow in a neighboring link.
This scenario is thus compatible with a stable front regime.
Using this argument, we can estimate the transition between the stable and columnar regimes by setting $h_c \sim 0$, which leads to the scaling relation:
\begin{equation}
    Ca_P \sim \frac{c R_\tau}{M}.
    \label{eq:Capc_column_stable}
\end{equation}
This scaling will be confirmed in the next section.

\section{Phase diagram} \label{sec:phase_diagram}

In this section, we summarize the different observed regimes as a function of the control parameters by locating them in a three-dimensional phase diagram.

\begin{figure*}[ht]
    \centering
    \includegraphics[width=0.7\hsize]{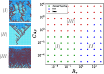}
    \caption{  Phase diagram for $M = 10^{-3}$. We observe the viscous fingering (I), directed tree (II) and foam (III) regimes. }
    \label{fig:phase_diagram_M0.001}
\end{figure*}

{\bf Low viscosity ratio:} Fig.~\ref{fig:phase_diagram_M0.001} presents the three distinct regimes identified for $M = 10^{-3}$ across a range of parameters.
We observe three transitional regimes.
At low $R_\tau$, there is a transition between viscous fingering and foam generation, which has been studied in~\cite{lanza23} for Newtonian fluids, and occurs at a critical capillary number $Ca_P$.
Moreover, according to this study, this critical value is expected to depend on the system size.

Another transition occurs at low $Ca_P$ between viscous fingering and the directed tree regime.
The vertical nature of this transition line suggests that it is governed by the competition between yield stress and capillary effects, which dictates the morphology of the invasion pattern.
The criterion used to determine the morphology is based on the tortuosity described in section \ref{sec:low_viscosity}.
We note that this criterion is not sufficiently sensitive to distinguish the "soft" transition between viscous fingering and invasion percolation.
Indeed, at very low $Ca_P$, the morphology appears to be slightly closer to invasion percolation. More generally, the transitions between regimes are relatively continuous, making the corresponding transition lines difficult to locate precisely. In particular, simulations performed near the boundary between two regimes often exhibit characteristics of both neighboring invasion patterns.
The last transition occurs at high $R_{\tau}$ and high $Ca_P$, where we observe a transition between the foam and the tree region. This occurs approximately at $Ca_P/R_{\tau} = \text{Cte}$, which suggests that it is governed by the competition between the viscous stress and the yield stress.
A possible interpretation would be as follows. Since it is the merging of finger tips that creates the foam \cite{lanza23}, foam generation requires a large number of Newtonian finger tips. On the other hand, the yield stress controls the number of tips, as described in Section~\ref{sec:low_viscosity}, through the ratio $Ca_P/R_{\tau}$.

\begin{figure*}[ht]
    \centering
    \includegraphics[width=0.7\hsize]{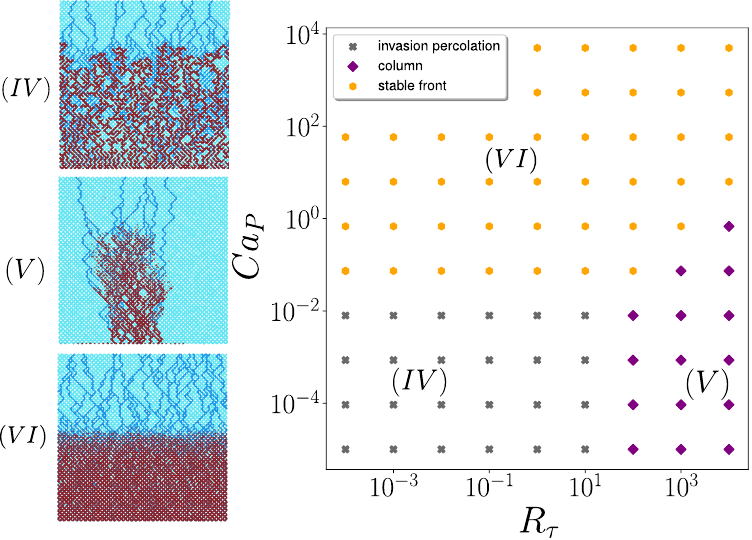}
    \caption{ Phase diagram for $M = 10^3$. We observe the stable front (IV), invasion percolation (V) and column (VI) regimes.
    }
    \label{fig:phase_diagram_M1000}
\end{figure*}

{\bf High viscosity ratio:}
 We now examine the phase diagram for a very high viscosity ratio ($M = 10^3$), as presented in Fig.~\ref{fig:phase_diagram_M1000}.
As discussed previously, three distinct regimes are observed: invasion percolation at low $Ca_P$ and low $R_\tau$, the column regime at low $Ca_P$ and high $R_\tau$, and a stable front at high $Ca_P$.
\begin{figure*}[ht]
    \centering
    \includegraphics[width=0.72\hsize]{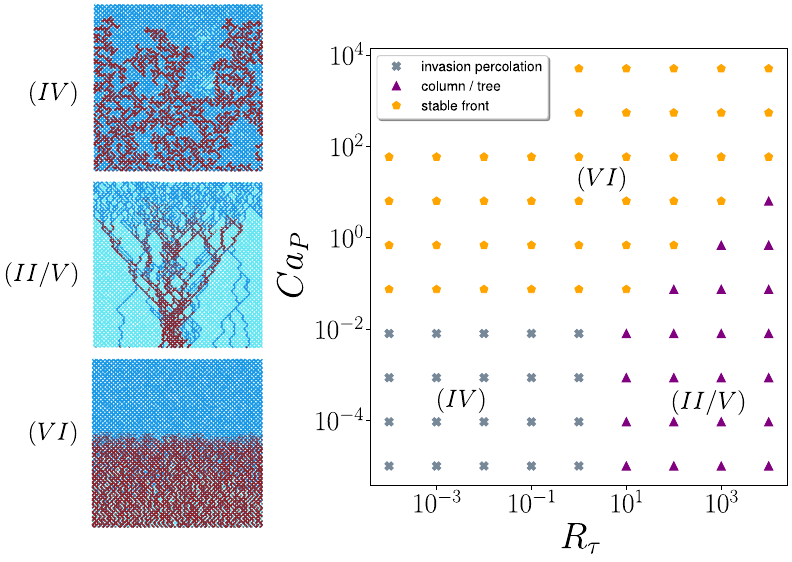}
    \caption{ Phase diagram for $M = 1$. We observe the invasion percolation regime (IV), an hybrid behavior (II/V) and a stable front regime (VI).
    }
    \label{fig:phase_diagram_M1}
\end{figure*}
The transition between the invasion percolation and column regimes occurs at a constant $R_\tau$, highlighting the competition between capillary forces and yield stress.
Although the transition between the stable front and invasion percolation regimes is relatively smooth, the transition line is horizontal, indicating that it occurs at a constant $Ca_P$ (in agreement with Lenormand diagram \cite{lenormand85}).
Finally, the transition between the stable front and the column regime occurs at a constant $R_\tau/Ca_P$ ratio, reflecting the competition between yield stress and viscosity.
Moreover, the estimation from Eq.~\eqref{eq:Capc_column_stable} in the previous section predicts a transition at $Ca_P \sim 10^{-3} R_\tau$, which aligns well with our simulation results.

{\bf Neutral viscosity ratio: }
The previous diagrams were constructed for extreme viscosity ratios, where the flow regimes are relatively well defined.
Here, we consider the case of equal viscosity for both fluids ($M = 1$), as shown in Fig.~\ref{fig:phase_diagram_M1}.
In this scenario, we still observe the stable front and invasion percolation regimes.
However, at high $R_\tau$, a hybrid flow regime emerges.
Initially resembling the column regime, with a primary path forming and widening at the bottom, this regime then transitions into directed branching at a certain distance from the inlet.
This behaviour arises from the combined effect of the two previously discussed mechanisms: the presence of a pressure drop within the column and an increasing pressure gradient in the yield-stress fluid region as the front propagates. Similar invasion patterns are observed when considering other intermediate values of the viscosity ratio, see Appendix~\ref{sec:diagram_intermediate} for the cases $M=10^{-1}$ and $M=10$. Finally, as shown in Appendix \ref{sec:rayleigh}, the invasion patterns reported here do not seem to depend on the specific distribution the radii of the network are sampled from.

\section*{Conclusion}
In this work, we have investigated the displacement patterns arising when a Newtonian fluid invades a porous medium saturated with an immiscible yield-stress fluid, by imposing a constant pressure drop between the inlet and the outlet. Our analysis reveals two previously uncharacterized flow regimes, governed by the dominance of yield stress over capillary and viscous effects.

At low viscosity ratios, a \textit{directed tree regime} emerges, where the invasion proceeds via successive branching from a primary path. This behavior stems from the assumption of nearly uniform pressure within the invading fluid, with the pattern dictated by the flow paths in the yield-stress fluid ahead of the front.
At high viscosity ratios, a \textit{column-like regime} appears, marked by progressive widening of the invasion path. Here, the mechanism is driven by viscous dissipation-induced pressure drops within the invading fluid, leading to the branching of new flow paths from the inlet. Using a homogeneous model, we derived a critical height for the invading path at which branching occurs.

We further explored the transitions between these new regimes and the established Newtonian regimes, viscous fingering and foam (low viscosity ratio), as well as stable front and invasion percolation (high viscosity ratio).  At intermediate viscosity ratios, hybrid regimes arise, where the front may initially resemble a column before transitioning into a tree-like structure.
Additionally, we established a breakthrough criterion for the Newtonian fluid as a function of system parameters.
To categorize these regimes, we introduced three dimensionless numbers: the capillary number $Ca_P$, the viscosity ratio $M$, and $R_\tau$, which quantifies the ratio of yield stress to capillary forces.

This study opens up a number of possibilities for further research. For example, fluid configuration could be extended to cases where a yield-stress fluid invades a Newtonian fluid or another yield-stress fluid. While we focused on drainage (non-wetting invading fluid), exploring imbibition (wetting invading fluid) would provide complementary insights. This study has been conducted using a constant pressure drop imposed at the inlet. Other inlet boundary conditions could be explored in future work. This is particularly important for non-Newtonian fluids, since the flow rate-pressure drop relationship cannot generally be inverted as for Newtonian flows. Consequently, different injection conditions may lead to different flow behaviors and possibly to different regimes \cite{talon24}.
Model refinements could incorporate additional physical mechanisms, such as shear-thinning or shear-thickening effects in the invading/defending fluid,  liquid films at the walls \cite{pourzahedi24}, or visco-elastic effects \cite{eslami17}. Investigating more realistic network topologies may also be important, since the exact shape of the invading structure in the tree regime may depend on the connectivity of the underlying network. Finally, experimental validation of these predictions would be valuable. Two-dimensional drainage experiments could, for example, be performed using a 3D-printed porous Hele-Shaw cell as the porous medium and a Carbopol-water solution as the Bingham fluid.

\section*{acknowledgements}
This work was partly supported by the Research Council of Norway through the INTPART program (project number 309139) and its Centers of Excellence funding scheme (project number 262644). AH also acknowledges funding from the European Research Council (Grant Agreement 101141323 AGIPORE).

\bibliography{biblio}

\appendix

\section{Validation of the critical pressure $\Delta P_0$} \label{sec:DP0_check}

In this appendix, we numerically verify that the critical pressure threshold $\Delta P_0$, defined in Eq.~\eqref{eq:DP0_def}, provides the correct criterion for the onset of breakthrough. To this end, we perform simulations for both limiting cases of the viscosity ratio $M$, over a range of values of $R_\tau$, and for negative values of the capillary number $Ca_P$, defined in Eq.~\eqref{eq:Ca_def}, ranging from $Ca_P=-0.005$ to $Ca_P=-0.1$ (the negative values are due to the fact we impose $\Delta P < \Delta P_0$).
The resulting invasion patterns are represented in the $(R_\tau,Bn_P)$ plane, where the Bingham number
\begin{equation}
    Bn_P = \frac{L \langle \tau \rangle}{\Delta P}
    = \frac{2\tau_cL/\langle r\rangle}{\Delta P}
\end{equation}
is recalled here for convenience. From the computed value of $\Delta P_0$, we define the corresponding critical Bingham number,
\begin{equation}
    Bn_{P,0} = \frac{L \langle \tau \rangle}{\Delta P_0},
\end{equation}
above which breakthrough is not expected to occur. It can be expressed in terms of the other dimensionless numbers as
\begin{equation}
    Bn_{P,0} = Bn_P - \frac{R_\tau}{Ca_P}.
\end{equation}

\begin{figure*}[h!]
    \centering
    \includegraphics[width=0.45\hsize]{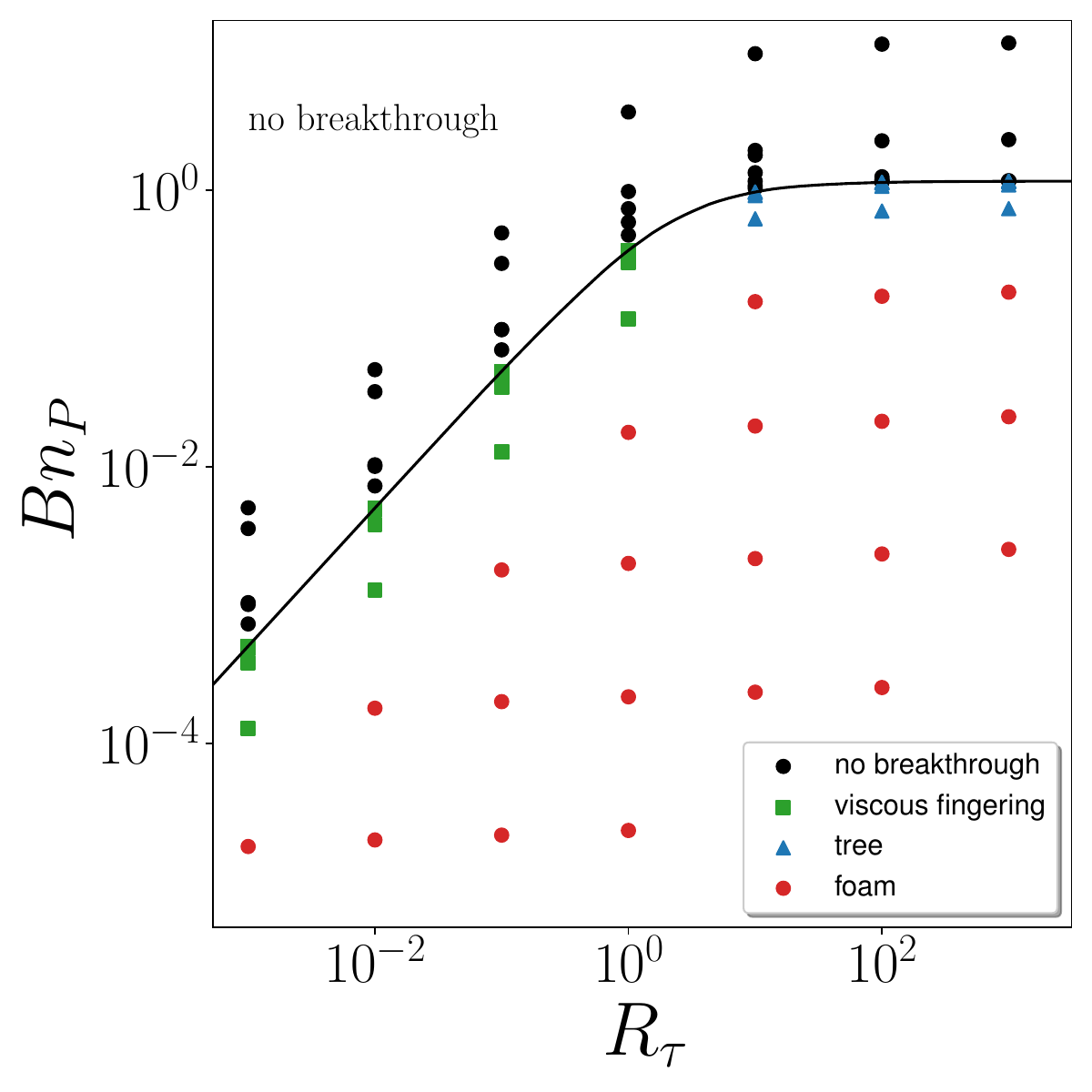}
    \includegraphics[width=0.45\hsize]{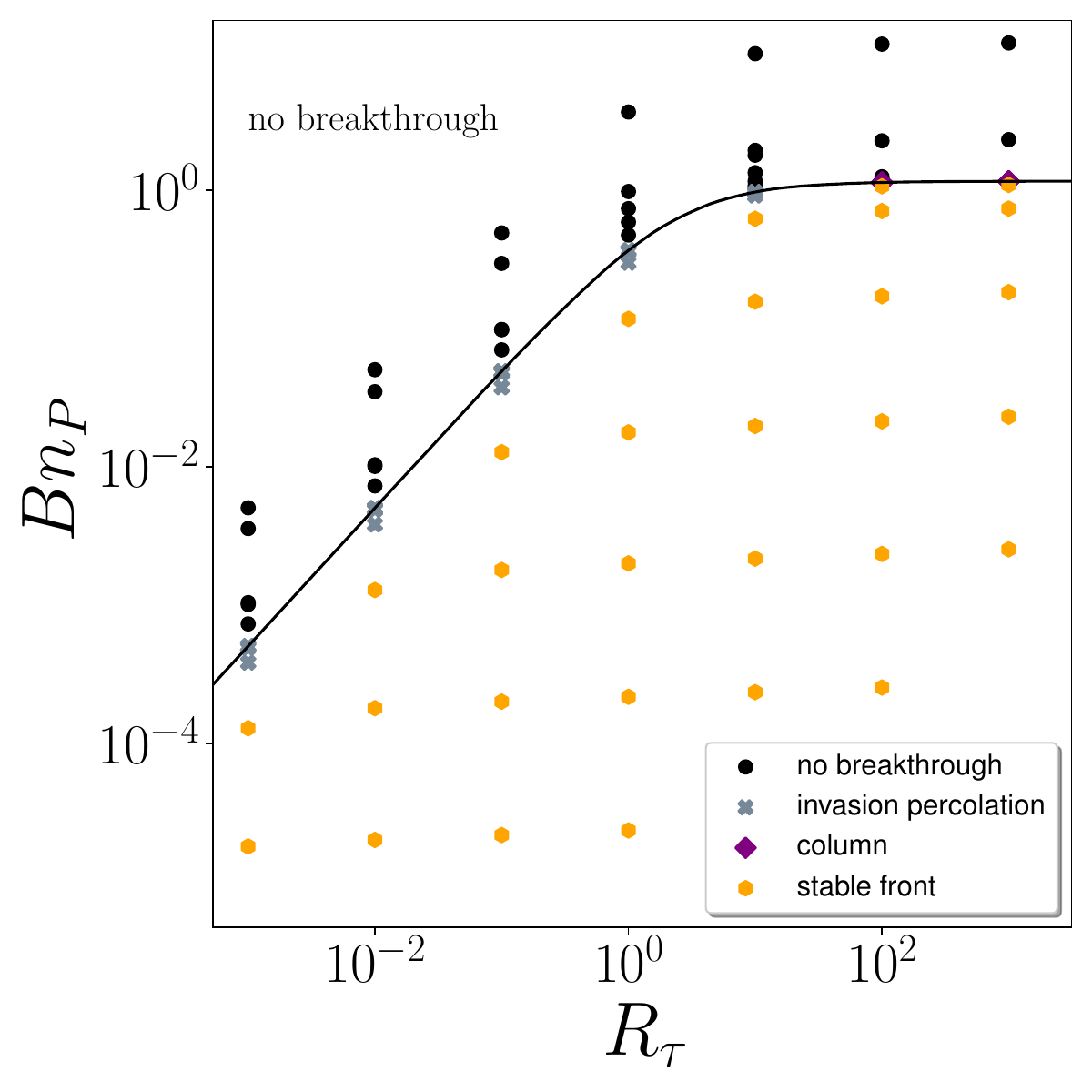}
    \caption{Phase diagram for $M = 10^{-3}$ (left) and $M = 10^{3}$ (right). The black solid line is the prediction for the critical Bingham $Bn_{P,0}$ using $\Delta P_0$, computed from Eq.~\eqref{eq:DP0_def}. The black dots represent the simulations where we imposed $\Delta P < \Delta P_0$, for which no breakthrough occurs. 
    }
    \label{fig:phase_diagram_DP0}
\end{figure*}

The results are presented in Fig.~\ref{fig:phase_diagram_DP0}. The critical Bingham number, shown as the solid black line, accurately separates breakthrough from no-breakthrough events over the entire range of parameters investigated, including cases lying very close to the predicted threshold.

\section{Phase diagrams for intermediate viscosity ratios $M$} \label{sec:diagram_intermediate}

\begin{figure*}[h!]
    \centering
    \includegraphics[width=0.75\hsize]{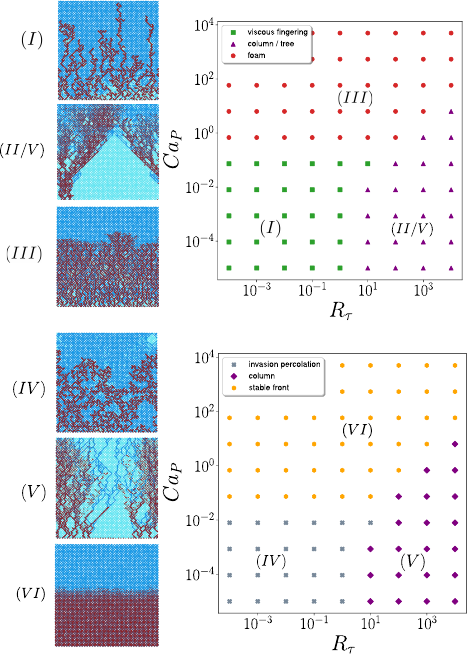}
    \caption{Phase diagram for $M = 10^{-1}$ (top) and $M=10$ (bottom).
    }
    \label{fig:phase_diagram_intermediate}
\end{figure*}

In this appendix, we briefly examine how the invasion patterns observed in the limiting cases of low and high viscosity ratios $M$ evolve for intermediate values of $M$. Figure~\ref{fig:phase_diagram_intermediate} shows the corresponding phase diagrams for $M=10^{-1}$ (left) and $M=10$ (right).
For $M=10^{-1}$, we observe essentially the same regimes as for the case $M=1$ discussed in Section~\ref{sec:phase_diagram}. The main difference with the $M=1$ diagram is the gradual transition from a stable-front regime towards a foam regime for the high $Ca_P$, low $R_\tau$ region. When yield-stress effects become dominant, we again observe the hybrid directed tree/column regime identified for $M=1$.

For $M=10$, the invasion patterns are qualitatively similar to those observed in the limiting case $M=10^3$. The main difference is that the column regime persists up to larger values of the capillary number $Ca_P$. This is consistent with the stabilizing effect of increasing the viscosity ratio, as discussed in Section~\ref{sec:high_viscosity}. However, the transition line predicted by Eq.~\eqref{eq:Capc_column_stable} no longer provides a quantitatively accurate description, overestimating the transition by approximately one order of magnitude. We attribute this discrepancy to the fact that, for intermediate viscosity ratios, the different invasion regimes are less clearly separated than in the asymptotic limits. The simplified arguments used to derive the transition criterion thus become less accurate.

\section{Rayleigh  distribution of radii} \label{sec:rayleigh}

\begin{figure*}[ht]
    \centering
    \includegraphics[width=0.75\hsize]{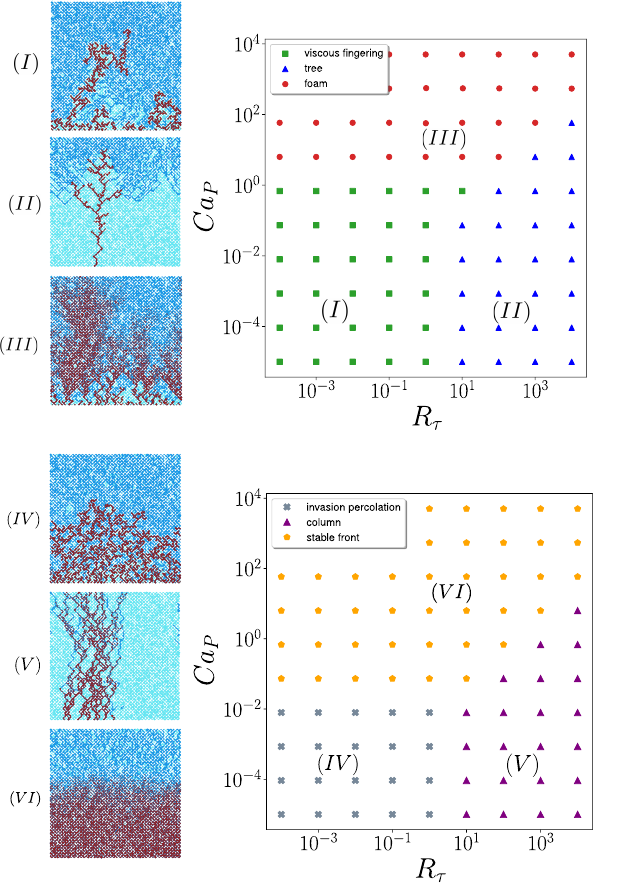}
    \caption{Phase diagram for $M = 10^{-3}$ (top) and $M=10^3$ (bottom) with a Rayleigh distribution of radii. The same displacement patterns are observed.
    }
    \label{fig:phase_diagrams_rayleigh}
\end{figure*}

In the main text, a uniform distribution for the radii was assumed, with values constrained between two strictly positive bounds. In this appendix, we investigate the sensitivity of the invasion patterns, presented in Section~\ref{sec:phase_diagram}, to the choice of radius distribution. Specifically, we assess whether the observed patterns are robust to alternative distributions by repeating our simulations using a Rayleigh distribution for the radii:
\begin{equation}
    \Pi(r) = \frac{r}{\sigma^2} \exp\left(-\frac{r^2}{2\sigma^2}\right),
\end{equation}
where $\sigma = 0.15$. Unlike the uniform distribution, the Rayleigh distribution permits radii arbitrarily close to zero, thereby introducing greater disorder into the medium. The resulting phase diagrams for $M = 10^{-3}$ and $M = 10^3$ are presented in Fig.~\ref{fig:phase_diagrams_rayleigh}.

Overall, the phase diagrams remain largely consistent with those obtained under the uniform distribution. However, subtle differences emerge in the boundaries between the observed regimes. Notably, the occurrence of column and tree regimes is more pronounced within their respective parameter spaces.
In the low-viscosity limit, this behavior can be attributed to the presence of very small radii in the medium, which increases the pressure cost associated with tortuous flow paths, thereby favoring the formation of directed trees. Additionally, the greater variability in the pressure threshold required to activate new flow paths inhibits the stabilization of a coherent front in the high-viscosity limit.


\end{document}